\PassOptionsToPackage{table}{xcolor}
\documentclass[sigconf]{acmart}

\usepackage{graphicx} 
\usepackage{amsmath}
\usepackage{enumitem}
\usepackage[ruled,shortend,linesnumbered]{algorithm2e}
\usepackage{multirow}
\usepackage{booktabs}
\usepackage{subcaption}
\usepackage{xcolor}
\usepackage{etoolbox}
\newtheorem{definition}{Definition}

\newtheorem{observation}{Inference}

\newcommand{\ie}{\textit{i.e.,} }
\newcommand{\eg}{\textit{e.g.,} }

\newcommand{\flash}{\texttt{ARX-LR}\xspace}
\newcommand{\arx}
{\texttt{ARX}\xspace}

\newcommand{\clinic}{\texttt{MAP}-clinics\xspace}

\newcommand{\colorcell}{\cellcolor[RGB]{192,192,192}\xspace}
\newcommand{\colorvalue}{\cellcolor[RGB]{203,255,204}\xspace}

\usepackage{tikz}
\newcommand*\circled[1]{\tikz[baseline=(char.base)]{\node[shape=circle,draw,inner sep=1.0pt] (char) {#1};}}

\renewcommand\footnotetextcopyrightpermission[1]{}

\begin{document}
\date{}
\fancyfoot[C]{\thepage}

\title{Exposing Privacy Risks in Anonymizing Clinical Data:\\ Combinatorial Refinement Attacks on \emph{k}-Anonymity Without Auxiliary Information}

\author{Somiya Chhillar}
\affiliation{
  \institution{George Mason University}
  \city{Fairfax, VA}
  \country{USA}
}
\email{schhilla@gmu.edu}

\author{Mary K.~Righi}
\affiliation{
  \institution{MAP Clinics}
  \city{Fairfax, VA}
  \country{USA}
}
\email{mrighi2@gmu.edu}

\author{Rebecca E.~Sutter}
\affiliation{
  \institution{George Mason University}
  \city{Fairfax, VA}
  \country{USA}
}
\email{rsutter2@gmu.edu}

\author{Evgenios~M.\ Kornaropoulos}
\affiliation{
  \institution{George Mason University}
  \city{Fairfax, VA}
  \country{USA}
}
\email{evgenios@gmu.edu}

\begin{abstract}
Despite longstanding criticism from the privacy community, $k$-anonymity remains a widely used standard for data anonymization, mainly due to its simplicity, regulatory alignment, and preservation of data utility. 
However, non-experts often defend $k$-anonymity on the grounds that, in the absence of auxiliary information, no known attacks can compromise its protections.

In this work, we refute this claim by introducing \emph{Combinatorial Refinement Attacks} (CRA), a new class of privacy attacks targeting $k$-anonymized datasets produced using local recoding. 
This is the first method that does not rely on external auxiliary information or assumptions about the underlying data distribution. 
CRA leverages the utility-optimizing behavior of local recoding anonymization of  \texttt{ARX}, which is a widely used open-source software for anonymizing data in clinical settings, to formulate a linear program that significantly reduces the space of plausible sensitive values. 
To validate our findings, we partnered with a network of free community health clinics, an environment where (1) auxiliary information is indeed hard to find due to the population they serve and (2) open-source $k$-anonymity solutions are attractive due to regulatory obligations and limited resources. 
Our results on real-world clinical microdata reveal that even in the absence of external information, established anonymization frameworks do not deliver the promised level of privacy, raising critical privacy concerns.
\end{abstract}

\maketitle

\pagestyle{empty}

\section{Introduction}

 Despite the criticism~\cite{DBLP:conf/innovations/CohenKMMNST25, DBLP:journals/pnas/CohenN20,10.1145/3452021.3458816,4529451,1617392,4221659,Altman2021Hybrid} by the privacy community, \emph{$k$-anonymity} remains a standard approach in practical anonymization and data privacy. 
Intuitively, $k$-anonymity formalizes privacy protection against re-identification by requiring that each individual’s data be indistinguishable from that of at least $k-1$ others, based on a specified set of quasi-identifiers (\ie sensitive attributes).
The popularity of $k$-anonymity stems from its (1) intuitive and accessible privacy guarantees, even for non-experts, (2) widespread availability of efficient open-source implementations, (3) preservation of data utility for statistical analysis and policy making compared to more rigorous privacy models, and (4) alignment with regulatory frameworks for data anonymization.

In this work, we focus on hierarchical $k$-anonymity for numerical sensitive attributes, where each value is generalized to a coarser interval that contains the original value. 
$k$-Anonymity comes in two flavors: \emph{global recoding}, where every equivalence class is formed using the same level of granularity; and \emph{local recoding}, where each equivalence class may use a different granularity level, provided that the resulting dataset is $k$-anonymous. 

The wave of linkage attacks on $k$-anonymity~\cite{Samarati2002kanonymity,1617392, DBLP:journals/corr/abs-0803-0032,10.1145/1993077.1993080} has not been sufficient to convince non-experts, who often counter that such attacks require sophisticated adversaries with access to auxiliary information (that is, information that allows the anonymized dataset to be linked with external data and uncover the identity of a participant). 
One could argue that without such auxiliary data, there is indeed little basis on which to mount any attack. 
Thus, the ongoing debate over the suitability of $k$-anonymity as an anonymization mechanism can be summarized as:

\begin{center}
\emph{``In the absence of external auxiliary information, a properly k-anonymized release provides strong protection.''}
\end{center}

A natural interpretation of the anonymized dataset is, in the absence of auxiliary information, the number of plausible values for a $k$-anonymized record is the product of the lengths of the generalized intervals across all sensitive attributes, \eg a generalized record with three attributes $([1$-$25],[50$-$100],[1$-$100])$ can be interpreted as any of the $25\cdot 50\cdot 100$ plausible underlying non-generalized records; and, if $k=4$ then, there are $(25\cdot 50\cdot 100)^4$ such interpretations for the entire equivalence class. 
Our work challenges this intuition by introducing a new class of attacks that drastically reduces the number of plausible interpretations \emph{without relying on any auxiliary information or prior distributional knowledge}.

\textbf{Combinatorial Refinement: The Price of Greed.} Our attacks target the local recoding variant of $k$-anonymity. 
The rationale behind targeting this mechanism is that local recoding preserves more information compared to global recoding; therefore, in settings where data analysts and policymakers prefer a more fine-grained view of the data while also satisfying $k$-anonymity, local recoding is preferred over its global recoding counterpart.
We term our approach \emph{Combinatorial Refinement Attacks} (CRA), as it departs from the standard assumption that all combinations within generalized intervals are equally plausible. Instead, our techniques systematically refine the space of plausible sensitive values by identifying feasible combinations across finer-grained subintervals, \ie only a fraction of the previously mentioned $(25 \cdot 50 \cdot 100)^4$ plausible underlying values are feasible given the observed anonymization.
At the core of our approach is the insight that local recoding algorithms, driven by greedy and optimal steps that maximize utility, leave behind patterns from their decisions that can be reverse-engineered to infer the presence (or absence) of sensitive records within subintervals.

\textbf{Differences from Downcoding Attacks~\cite{DBLP:conf/uss/Cohen22}.} A recent work by Aloni~\cite{DBLP:conf/uss/Cohen22} introduces downcoding attacks on $k$-anonymization mechanisms. Although~\cite{DBLP:conf/uss/Cohen22}  claims that these attacks do not require auxiliary information, distributional knowledge of the data is still necessary. 
Specifically, Theorems 4.2 and 4.3 in~\cite{DBLP:conf/uss/Cohen22} demonstrate that \emph{there exists} a data distribution and a generalization hierarchy under which downcoding is possible. 
However, the existence of such a (potentially contrived) instantiation does not imply that the attack is broadly applicable or effective in real-world scenarios. 
As such, the practical relevance and generalizability of downcoding attacks remain unclear.
More importantly, while the downcoding attacks in~\cite{DBLP:conf/uss/Cohen22} are claimed to operate without access to non-generalized records (\ie without auxiliary information), they do assume full knowledge of the underlying data distribution. In fact, the attack logic is explicitly tailored to the specific characteristics of the data distribution under attack. 
As a result, the applicability of this attack is limited to the particular scenario presented in the paper. 
On the contrary, our combinatorial refinement attacks can be applied to any locally recoded dataset without any information about the data distribution and without any auxiliary information.

\textbf{A Real-World Scenario: Anonymizing Clinical Data.} The motivation of this project came from the collaboration of our team with \clinic, an interprofessional network of community health clinics affiliated with George Mason University that provides services to uninsured and refugee populations in underserved areas of a metropolitan region. 
Operating under a bridge-care model, these free clinics offer a range of services, including primary health care, school physicals, screenings, and mental health support to individuals in low-income and medically underserved communities.
\clinic holds access to microdata that can significantly inform policy-making decisions in the region; as such, access to this data is of paramount importance for both policymakers and researchers. 
Given that such clinics across the country often operate under tight budget constraints, it is reasonable for them to adopt non-patented anonymization method,s particularly those available through open-source codebases and already trusted by peer institutions.

Our team reviewed the literature to help \clinic explore suitable tools for potential future adoption. 
Our research showed that \texttt{ARX}~\cite{prasser2014arx} is widely recognized as the leading open-source tool~\cite{ARXTool} for $k$-anonymization, particularly in healthcare~\cite{prasser2014arx,10.1093/bib/bbac440, Pilgram2024Costs,fi14060167}.
It is a Java-based platform offering both a user-friendly GUI and a well-documented API. 
\texttt{ARX} is praised for its robust risk and utility analysis features, regulatory compliance (e.g., HIPAA, GDPR), and ease of use for both technical and non-technical users. 
Actively maintained since 2012, it has seen broad adoption across commercial platforms, research projects~\cite{Meurers2021Scalable, Jakob2020Anonymization, Eicher2020PredictionTool,10.1093/ehjdh/ztae083,Bild2020BetterSafe,Spengler2019AttributeDisclosure,Bild2020SensitiveAttribute}, and clinical trials 
~\cite{Pilgram2024Costs, Im2024PrivacyUtility}.
Beyond these academic projects, \texttt{ARX} has also been used in real-world healthcare settings. 
Researchers at the Cancer Registry of Norway selected \texttt{ARX} as one of their preferred de-identification technique for processing over 5 million health records from the Norwegian Cervical Cancer Screening Program~\cite{Ursin2017Fuzzy}. 
Additionally, \texttt{ARX} has been recognized in official policies and guidelines as a recommended tool for anonymizing biomedical data, e.g., the UK Anonymisation Network  \cite{EMA2025} and the European Medicines Agency~\cite{Elliot2016} responsible for the scientific evaluation, supervision, and monitoring of medicines for human and veterinary use.
One of the algorithms for $k$-anonymity is called \texttt{FLASH}, and it is a global-recoding algorithm that was introduced in~\cite{DBLP:conf/socialcom/KohlmayerPEKK12}.
 Our privacy assessment focuses on the local recoding variant of \texttt{ARX} that uses \texttt{FLASH} iteratively to carve out optimal equivalence classes from the remaining dataset; in the rest of this work, we refer to this combination from \texttt{ARX}~\cite{arx-github} as \flash. 
Interestingly, our technique is effective regardless of which globally optimal algorithm is used in place of \texttt{FLASH} to construct a local recoding mechanism; that is, our findings are not specific to the design of \texttt{FLASH} itself.

\textbf{Our Contributions.} Our contributions are:
\begin{itemize}[leftmargin=*,topsep=0pt]
\item We revisit the open-source codebase of the local recoding algorithm \flash and present a simplified but functionally equivalent version in Section~\ref{sec:flash}.
\item We analyze how the utility-driven greedy decisions made by \flash reveal information about subsequently formed equivalence classes. 
 Building on this insight, we develop a series of inferences (detailed in Section~\ref{sec:greedy}) regarding data's location during \flash’s execution. 
 We then formalize these observations into a Combinatorial Refinement Attack by introducing a linear programming formulation. 
 While the objective function is indifferent, the linear constraints encode the inferences uncovered during \flash’s execution. 
 We then enumerate all feasible integer solutions to this program, which gives a refined set of plausible non-generalized records for the target equivalence class.
 \item Given the potential scalability limitations of integer programming and solution enumeration, we assess the practical viability of the proposed CRA on real-world data. 
 We partner with a network of free \clinic and test CRA on micordata anonymized by \flash. 
 We also test CRA on anonymized microdata from the Healthcare Cost and Utilization Project (HCUP). 
 Across both datasets, we identified anonymized equivalence classes where the application of CRA reduced the set of plausible records by $7$-$39,000\times$ on average, relative to the number implied by \flash.
\end{itemize}

\noindent\textbf{Ethical Considerations.} 
The study was conducted in coordination with our IRB office. 
All reported statistics are aggregated and anonymized; no PII, raw values, or statistics about the data are disclosed in this manuscript.

\noindent\textbf{Vulnerability Disclosure.} We disclosed our findings about the existence and the effectiveness of combinatorial refinement attacks to the developers of \texttt{ARX} on April 10th, 2025.
The \texttt{ARX} team confirmed receipt on the same day. On August 9, they updated the documentation to warn about the local recording algorithm’s susceptibility to inference attacks.
\section{Preliminaries}

In this section, we introduce standard terminology from the anonymization literature  (\eg generalization, $k$-anonymity, generalization hierarchy, generalization lattice, global/local recoding) as well as newly introduced terminology (such as basic and compound segments) that we will use in the rest of the work. 

\textbf{Notation and Terminology.} Let $A = (A^{1},..,A^{m})$ be the $m$ dimensional space of attributes. 
The notation $[\alpha,\beta]$ denotes an interval that includes all values between $\alpha$ and $\beta$, inclusive, while notation $[\alpha,\beta)$ (resp. $(\alpha,\beta]$) does not contain the last (resp. first) value of the interval.
The \emph{attribute domain} $\emph{Dom}(A^{i})$ of an attribute $A^{i} \in A$, where $i \in [1,m]$, represents the set of all possible values for attribute $A^{i}$. 
In this work, we focus on \emph{numerical attributes}, as opposed to categorical attributes.
A \textbf{\emph{data record}} (or simply record), $x = (x^{1},\ldots,x^{m})$ is an $m$-dimensional vector where each attribute $A^{i}$ takes a single value $x^{i}$ from its domain $\emph{Dom}(A^{i})$. 
A \emph{dataset} $D$ is a collection of data records, and its cardinality is denoted as $n=|D|$.
\emph{Quasi-identifiers} $Q$ are a subset of privacy-sensitive attributes that, if they appear in a public dataset, they can be used to enable linkage attacks.
In this work, we take a privacy-conservative approach and consider \emph{all} attributes to be quasi-identifiers, meaning that each record is a vector of quasi-identifiers, \ie $(Q^{1},..,Q^{m})$. 
For simplicity, we assume that all attribute values are distinct within each domain~\footnote{If identical values are allowed across records, it becomes possible to ``group'' them without applying generalization through intervals; an approach that is technically $k$-anonymous but, in our view, overly revealing and, thus, we avoid it.}, \ie no two records share the same value for any quasi-identifier.

To break the tension between privacy and utility, the community has studied techniques that transform data records,  known as \emph{generalizations}, in which a more coarse-grained representation replaces each value. 
This way, some statistical properties are preserved (maintain utility), while anonymity is seemingly preserved. 
A \textbf{\emph{generalized record}} $y = (y^{1},\ldots,y^{m})$ is an $m$-dimensional vector where each value $x^{i}$ is replaced by an \emph{interval} of consecutive values $y^{i}$ from the corresponding domain $\emph{Dom}(Q^{i})$, where $i \in [1,m]$. 
Formally, $y = (y^{1},\ldots,y^{m})$ generalizes  $x = (x^{1},\ldots,x^{m})$ if for all $i\in[m]$ we have that $x^i\in y^i$. 
Intuitively, a generalized record ``hides'' the true value of each quasi-identifier by only indicating an interval of the domain to which the value belongs. 
For example, the record $x = (95,23)$ with quasi-identifiers $Q^{1} = \texttt{Blood Glucose Level}$ and $Q^{2} = \texttt{Age}$,  can be generalized to $y = ([75,100], [1,25])$.

\textbf{$k$-Anonymity.} In this work, we only focus on hierarchical $k$-anonymity. On a high-level a dataset satisfies \emph{k-anonymity} if (1) every generalized record $y$ appears at least $k$ times within the dataset or (2) if it does not appear $k$ times, then the generalization contains the interval that spans the entire domain of each attribute, also known as outliers.
The set of identical (and therefore indistinguishable) records is called an \emph{equivalence class (EQ)}. 
If a record is generalized to contain the entire domain for all quasi-identifiers (that would be the case of an outlier) it is said to be \emph{suppressed}, \ie $y=(\emph{Dom}(Q^{1}),\emph{Dom}(Q^{2}))$.
For a detailed definitional treatment, we refer the reader to~\cite{samarati1998generalizing, Samarati2002kanonymity, 10.1142/S021848850200165X}.

\begin{figure*}[t] 
    \centering
    \includegraphics[scale=0.258]{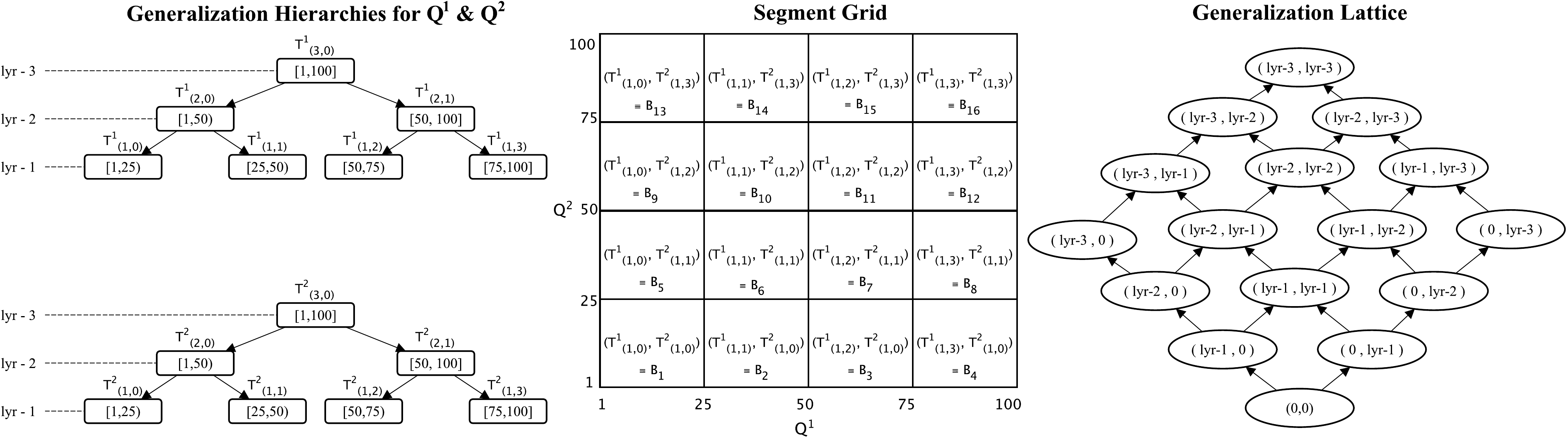}
    (a) \hspace{5cm} (b) \hspace{5cm} (c)
    \vspace{-0.3cm}
    \caption{(a) The generalization hierarchies define the possible granularity levels for partitioning the domain of each quasi-identifier.
(b) The segment grid provides a geometric representation in the $m$-dimensional space, illustrating the simultaneous selection of partition sets within a given generalization hierarchy.
(c) The generalization lattice captures the granularity options in terms of layers within each generalization hierarchy.}
    \label{fig:prelim}
\end{figure*}

\textbf{Generalization Hierarchy.} 
A core concept of $k$-anonymity is the notion of a generalization hierarchy for quasi-identifier $Q^{i}$, which defines a structured set of transformations that partition $\emph{Dom}(Q^{i})$ at varying degrees of granularity.
Figure 1 illustrates an example of generalization hierarchies for two quasi-identifiers $Q^1$ and $Q^2$ that (for simplicity) have the same domain $\emph{Dom}(Q^{1})=\emph{Dom}(Q^{2})=[1,100]$.
 More formally, the \textbf{generalization hierarchy} $T^{i}$ of a quasi-identifier $Q^{i}$ is a rooted tree of height $h^i$, denoted as $T^{i}=(T^{i}_{h^i},\ldots,T^{i}_{1})$ with $h^i$ layers. 
 We draw attention here to the (rather counterintuitive) convention that the root/top node of the tree sits at layer $h^i$, while the leaves reside at layer $1$, \ie the reverse of the standard terminology for trees in data structures. Jumping ahead, this convention is chosen so that the tree notation aligns with the standard in the literature description of generalization lattices. 
 In general, the $lyr^i$-th layer $T^{i}_{{lyr^i}}$ of the $i$-th hierarchy $T^{i}$ comprises a partition of the domain $\emph{Dom}(Q^{i})$ where each interval of the partition has the same length.
 The first interval\footnote{The $h^i$-th layer has only a single interval $T^{i}_{(h^i,0)}$} of the $j$-th layer is denoted as $T^{i}_{(j,0)}$ while the last as  $T^{i}_{(j,2^{(h^i-j)}-1)}$. 
 We use the notation $T_{(lyr^i,r^i)}^i$ to refer to each node in the generalization hierarchy $T^i$, where $lyr^i$ represents the layer of the node in the tree, and $r^i$ denotes its rank, indicating its position within that layer $lyr^i$, indexed from left to right.
Each node of a generalization hierarchy $T^{i}$ represents a sub-interval $[\alpha,\beta)$ of $\emph{Dom}(Q^{i})$, and all direct children of this node collectively form a further partition of $[\alpha,\beta)$.
For example, in Figure~\ref{fig:prelim}, the node corresponding to the interval $[1,50)$, is partitioned into the intervals $[1,25)$ and $[25,50)$ by its children nodes.
We note here that in this work, we analyze generalization hierarchies that are binary trees, but our approach can be easily extended to trees with any constant fanout.

\textbf{Generalization State.} 
The generalization of a quasi-identifier $Q^i$ is guided by the options in the corresponding generalization hierarchy $T^i$. The specific degree of generalization applied to a record $x$ is determined by its generalization state. 
The generalization state consists of a vector of layers, one for each quasi-identifier from the corresponding $T^i$, which determines the level of granularity applied to the generalization of each quasi-identifier in $x$. 
Formally, the \emph{generalization state}
is represented as a tuple $g = (lyr^1, \dots, lyr^m)$, where $lyr^i$ represents a layer from $1$ to $h^i$ of hierarchy $T^i$. 
The generalization state dictates the partition that will be used from each quasi-identifier to transform $x$ to $y$. 
When a quasi-identifier is not generalized at all, \ie instead of a range we have a single value, then we say that the layer for this quasi-identifier is $0$. See the nodes in Figure~\ref{fig:prelim}(c) for an illustration of a generalization state. 

The term $g(D)$ denotes the result of applying the generalization state $g$ to the dataset $D$. The notation $g(D).Q^i$ denotes the vector of intervals of $g(D)$ when considering only the quasi-identifier $Q^i$. 

\textbf{Generalization Lattice. } A \emph{generalization lattice} $L$ is a partially ordered set of generalization states that provides the search space for forming equivalence classes via a $k$-anonymization algorithm.
An example of a generalization lattice for a dataset with quasi-identifiers $Q^1, Q^2$ is shown in Figure~\ref{fig:prelim}(c).
The bottom-most node represents the original record with no generalization, \ie the node is the generalization state $g = (0,\ldots,0)$. 
The top-most node corresponds to the maximum possible generalization, \ie $g =(h^1,\ldots,h^m)$.
In our example, the maximum generalization state for $Q^1$ and $Q^2$ is $3$.
Edges in the lattice represent a transition in which exactly one quasi-identifier is generalized further by ``going up'' one layer in the tree. 

\textbf{Global and Local Recoding.} There are two main approaches for making a dataset $k$-anonymous. The first one is called \emph{global recoding}~\cite{Samarati2002kanonymity}, and in this approach, all equivalence classes belong to the same generalization state. 
Specifically, \textbf{every data record adopts the same generalization state}, \ie $T^i_{lyr^i}$ at layer $lyr^i$ for quasi-identifier $Q^i$, where $i\in[1,m]$. 
This approach is suboptimal with respect to the richness of information given to data analysts (\ie utility) since there might be some equivalence classes that could have been generalized ``less'' while still being $k$-anonymous. Which leads us to the second approach called \emph{local recoding}~\cite{10.1145/775047.775089} in which \textbf{each equivalence class can be generalized with respect to a different generalization state}.

\textbf{Segment.} For our analysis, we introduce the term \emph{segment}. 
Segments provide a geometric re-interpretation of the simultaneous selection of $m$  nodes across generalization hierarchies.
We define two types of segments: basic segments and compound segments.
A \textbf{\emph{basic segment}} comprises $m$ \emph{leaf nodes}, one from each of the $m$ generalization hierarchies. 
Essentially, this is the most revealing generalization interval since for each dimension/quasi-identifier, we take the most fine-grained generalization that is represented by a leaf.
More formally, a \emph{basic segment} is defined as the set 
\begin{displaymath}
B = \left(T^1_{(1, r^1)},\ldots,T^m_{(1, r^m)}\right)\text{,  where } r^i \in [0,2^{(h^i-1)}-1].
\end{displaymath}

Notice that each tree contributes a leaf node, at layer 1, across all $i$.
A \textbf{\emph{compound segment}} represents a ``higher level'' of generalization in which not all quasi-identifiers remain at the leaf layer; instead, some are generalized to a ``coarser granularity''. The term compound signifies that at least one quasi-identifier is generalized to include \emph{multiple basic segments}, making it an internal node within its corresponding generalization hierarchy.
Formally, a \emph{compound segment} is defined as 
\begin{displaymath}
\begin{split}
C = \left(T^1_{(lyr^1, r^1)},\ldots,T^m_{(lyr^m, r^m)}\right)\text{,  s.t. } r^i \in [0,2^{(h^i-1)}-1], lyr^i \geq 1,\\\text{and there is an entry s.t. } lyr^i \neq 1 .
\end{split}
\end{displaymath}
We note that the last requirement of the set $C$ guarantees that not all quasi-identifiers are leaves.
The term \emph{segment grid}, see Figure~\ref{fig:prelim}(b), illustrates how basic and compound segments are structured. 
Each cell in the grid represents a basic segment $B_i$, consisting of a pair of nodes $(T^1_{(lyr,r)}, T^2_{(lyr',r')})$ from two generalization hierarchies of $Q^1$ and $Q^2$, respectively. 
Individual cells labeled $B_1, B_2, \dots, B_{16}$ correspond to basic segments. 
We start the numbering from the bottom-left corner and go row-wise.
For example, in Figure~\ref{fig:prelim}(b), consider the generalization state $g=(lyr_3,lyr_1)$ which implies the following generalization options $([1,100], [1,25))$, $([1,100], [25,50))$, $([1,100], [50,75))$, and $([1,100], [75,100))$.
This generalization corresponds to four compound segments and each of them can be expressed as the union\footnote{Here, we abuse notation and assume that the union of consecutive intervals in the $m$-dimensional space is itself an interval.} of basic segments, \ie compound segments $\{B_1 \cup B_{5} \cup B_{9} \cup B_{13}\}$, $\{B_2 \cup B_{6} \cup B_{10} \cup B_{14}\}$, $\{B_3 \cup B_{7} \cup B_{11} \cup B_{15}\}$, and $\{B_4 \cup B_{8} \cup B_{12} \cup B_{16}\}$.
The basic segments that appear in the union that forms a compound segment are said to be contained within the compound segment. More formally:

\begin{definition}
Let $m$ be the number of quasi-identifiers in the dataset.
Let $S_A=\left(T^1_{(l_A^1, r_A^1)},\ldots,T^m_{(l_A^m, r_A^m)}\right)$ be a segment (basic or compound), where $T^i_{(l_A^i, r_A^i)}$ is a node in the generalization tree $T^i$ for quasi-identifier $Q^i$.
Let $S_B=\left(T^1_{(l_B^1, r_B^1)},\ldots,T^m_{(l_B^m, r_B^m)}\right)$ be a compound segment, where $T^i_{(l_B^i, r_B^i)}$ is a node in the generalization tree $T^i$ for quasi-identifier $Q^i$.
We say segment $S_A$ is \textbf{contained} in (compound) segment $S_B$, denoted as $S_A \in S_B$, if for every quasi-identifier $Q^i$, the node $T^i_{(l_A^i, r_A^i)}$ is a decendent of (or equal to) $T^i_{(l_B^i, r_B^i)}$ in $T^i$.
\end{definition}
We emphasize here that not every possible set of basic segments defines a compound segment. 
Compound segments are required to correspond to a valid generalization state. 
This is possible only when the set of leaf nodes representing the basic segment is a descendant of the set of nodes representing a compound segment for each quasi-identifier. 
For example, $([1,25),[51,75))$ cannot be a basic segment contained in the compound segment $([1,25),[1,50))$ because $[51,75)$ is not a descendant of $[1,50)$.
 
We say that a (non-generalized) record $x = (x^1,\ldots,x^m)$ \emph{belongs to segment} $C$ (compound or basic) if  $\forall i\in [1,m]$, we have that $x^i$ belongs to the interval of $T_{(lyr^i,r^i)}^i\in C$. 
If $\lvert C \rvert \geq k$ (or 
$\lvert B \rvert \geq k$), then the segment is an equivalence class.

\section{The \flash Algorithm for Local Recoding}
\label{sec:flash}

In this section, we restate the \flash algorithm for local recoding from the codebase of \texttt{ARX}~\cite{arx-github}. 
A key advantage of \flash is that it is open-sourced, making it accessible, auditable, and widely adopted,  particularly in academic research, including applications in the clinical domain. 
These qualities make \flash an attractive option for smaller organizations that lack the resources to to access proprietary anonymization tools.
This accessibility and practical relevance are central to our decision to focus on breaking the definition of k-anonymity for \flash in this work. 

\textbf{Information Loss.} The goal of $k$-anonymization is to resolve the tension between privacy and utility in data anonymization. 
On one hand, $k$-anonymization  enhances privacy by generalizing records, making the data more ``coarse-grained''. 
On the other hand, this generalization also reduces the amount of retained information. From an analyst’s perspective, finer-grained generalization of data preserves more details, resulting in higher utility.
A key metric for quantifying this tradeoff is \emph{information loss}, which measures the extent to which data utility is reduced due to anonymization. 
Intuitively, the information loss for a quasi-identifier is \emph{minimized} when it is not generalized at all, \ie the value of the record remains at layer $0$ of its generalization hierarchy. 
Conversely, the information loss for a quasi-identifier is \emph{maximized} when the quasi-identifier is generalized to the highest possible layer (root) of its hierarchy, \ie the layer is $h^i$, and represents the entire domain.
More formally, to measure a quasi-identifier's information loss, denoted as $Q\_loss^i$, at layer $lyr^i$ of its corresponding generalization hierarchy, one of the options the local recoding \flash  uses is the following formula: 
\begin{equation}
Q\_loss_{lyr^i}^i = \frac{\text{interval\_length}\left(T^i_{(lyr^i,*)}\right)}{\text{interval\_length}\left(T^i_{(h^i,0)}\right)}\cdot n,
\label{formula:QI_loss}
\end{equation}
where the term $T^i_{(lyr^i,*)}$ indicates any partition set from layer $lyr^i$ of $T^i$. 
We note here that $T^i_{(h^i,0)}$ is the interval that contains the entire domain. 
Term $Q\_loss_{lyr^i}^i$ takes the minimum value when the layer is $lyr^i=0$; hence the numerator is $1$, \ie $min^{(i)}\triangleq Q\_loss_0^i$. 
And maximum when the quasi-identifier represents the entire domain,  \ie $max^{(i)}\triangleq Q\_loss_{h^i}^i=n$. 
The \flash algorithm normalizes the loss of each quasi-identifier by:
$Q\_loss_{lyr^i}^i = (Q\_loss_{lyr^i}^i - min^{(i)})/(max^{(i)} - min^{(i)})$. 
Finally, one of the metrics provided by \texttt{ARX} is the total information loss, denoted by $\text{loss}_g$, for a generalization state $g = (lyr^1, \dots, lyr^m)$. 
It is computed as the geometric mean of the individual information losses across all quasi-identifiers: 
\begin{equation}
loss_g = \left(\prod_{i=1}^{m} \left(Q\_loss_{lyr^i}^i  + 1\right)^{1/m}\right) -1.
\label{formula:loss}
\end{equation}

Looking ahead, \flash evaluates the generalization loss $loss_g$ across all nodes in the lattice to identify a subset of $D$ on which $g$ can be applied such that the result generalizes to one or more equivalence classes, each containing at least $k$ records.

\textbf{\flash Criteria.} 
The \flash local recoding algorithm traverses the generalization lattice to identify a generalization state that satisfies two conditions: ($i$) when applied to the dataset $D$, and it forms at least one $k$-anonymous equivalence class, and ($ii$) among all generalization states meeting condition ($i$), it chooses the one with the \emph{minimum information loss}. 
However, a situation may arise where multiple 
generalization states have the same information loss and each of them is a candidate to form an equivalence class of size $k$.
To resolve such ties, \flash uses three tie-breaking \emph{criteria}. 
The first criterion $c_1(g)$ is a numerical value that captures the generalization across all quasi-identifiers in $g$ by adding the layer values, see equation~(\ref{formula:criteria}). 
The second criterion $c_2(g)$ normalizes the value of each individual layer by dividing it by the height of the tree, see equation~(\ref{formula:criteria}). 

The third criterion $c_3$ is a function not only of $g$ but also of $D$. 
In the denominator of each term in $c_3$, we count the number of distinct values in $Q^i$   in the non-generalized $D$, which we denote as $dst(D.Q^i)$. 
In the numerator of each term in $c_3$, we count the number of distinct intervals in $Q^i$  that result from applying the generalization $g$ to the $Q^i$ attribute of the records in $D$, which we denote as $dst(g(D).Q^i)$. 
Let $g=(lyr^1,\ldots,lyr^m)$ the criteria are:
\begin{equation} \label{formula:criteria}
\begin{aligned}
c_1(g) = \sum_{i=1}^{m} lyr^i \hspace{0.5cm}&
c_2(g) = \frac{1}{m} \cdot \sum_{i=1}^{m} \frac{lyr^i}{h^i} \\
c_3(g, D) &= 1 - \frac{1}{m} \cdot \sum_{i=1}^{m} \frac{dst(g(D).Q^i)}{dst(D.Q^i)}
\end{aligned}
\end{equation}

These criteria are evaluated in sequence $c_1$, $c_2$, and finally $c_3$.
If two generalization states have identical information loss, the algorithm picks the one with the lower $c_1$ value.
If the first criterion is not enough to resolve the tie, then the algorithm compares $c_2$.
The generalization state with lower $c_2$ is selected.
Finally, if both $c_1$ and $c_2$ are identical, then the algorithm resorts to $c_3$ to break the tie.
The generalization state with the lower $c_3$ value is picked.

\begin{algorithm}[h!]
    \small
    \caption{\texttt{\flash-Anonymizer}\label{algo:anonymizer}}
    \KwData{A non-generalized dataset $D$, an anonymity parameter $k$, and a generalization lattice $L$}
    \KwResult{A generalization state}
    
    Initialize \texttt{optimal\_state} as NULL.\;
\For{each generalization state $g_{\text{current}}$ in the generalization lattice $L$}{
     \If{$k$-anonymous equivalence classes are formed when applying $g_{\text{current}}$ to $D$}{
             \uIf{\texttt{optimal\_state} is NULL}{
                Assign $g_{\text{current}}$ to \texttt{optimal\_state}.\;
             }
              \uElseIf{$loss_{g_{\text{current}}} < loss_{\text{\texttt{optimal\_state}}}$}{
                      Assign $g_{\text{current}}$ to \texttt{optimal\_state}.\;        
              }\ElseIf{$loss_{g_{\text{current}}} = loss_{\text{\texttt{optimal\_state}}}$}{
                    \uIf{Criterion $c_1$ of $g_{\text{current}}$ is lower than $c_1$ of \texttt{optimal\_state}}{
                        Assign $g_{\text{current}}$ to \texttt{optimal\_state}.\;
                    }
                    \uElseIf{Criterion $c_1$ values of $g_{\text{current}}$ and  \texttt{optimal\_state} are identical but criterion $c_2$ of $g_{\text{current}}$ is lower than $c_2$ of \texttt{optimal\_state}}
                    {
                        Assign $g_{\text{current}}$ to \texttt{optimal\_state}.\;
                    }
                    \ElseIf{Criterion $c_1$ and $c_2$ values of $g_{\text{current}}$ and  \texttt{optimal\_state} are identical but criterion $c_3$ of $g_{\text{current}}$ is lower than $c_3$ of \texttt{optimal\_state} }
                    {
                        Assign $g_{\text{current}}$ to \texttt{optimal\_state}.\;
                    }
              
              }

    }
}
Once all states have been evaluated, return \texttt{optimal\_state}\;
\end{algorithm}
\begin{algorithm}[h!]
        \small
        \caption{\texttt{\flash-LocalRecoding}\label{algo:flash}}
        \KwData{A non-generalized dataset $D$, an anonymity parameter $k$}
        \KwResult{A locally recorded dataset}
        Generate the generalization lattice and store it in $L$ \;
        Store a copy of $D$ in \texttt{updated\_data}. \;
        Initialize \texttt{optimal\_state} as NULL \;
        Initialize \texttt{locally\_recoded} as NULL\;
        \While {there are at least $k$ records in \texttt{updated\_data}}
        {
            \texttt{optimal\_state} $\gets$ Anonymizer(\texttt{updated\_data}, $k$, $L$) \;
            \uIf{\texttt{optimal\_state} $== (h^1,\ldots,h^m)$}
            {
                break out of the loop. \;
            }
            \Else
            {
                Apply \texttt{optimal\_state} to \texttt{updated\_data} \;
                Store the $k$-anonymous records to \texttt{locally\_recoded} resulted from \texttt{optimal\_state} application\;
                Update \texttt{updated\_data} to contain the remaining non-anonymous records\;
            }
        }
        Suppress and append remaining records in \texttt{updated\_data} as outliers to \texttt{locally\_recoded} \;
        Return \texttt{locally\_recoded} \;
\end{algorithm}

\textbf{The \flash Algorithm.} \flash is presented in two subroutines, the anonymizer (which is Algorithm~\ref{algo:anonymizer}) and the local recoding (which is Algorithm~\ref{algo:flash}). 
We present an equivalent, though computationally less efficient, variant of the core algorithm \texttt{FLASH} proposed in~\cite{DBLP:conf/socialcom/KohlmayerPEKK12}. 
While the original algorithm is optimized for performance, both versions ultimately yield the same output. 
At a high level, our version exhaustively explores each generalization $g$ in the generalization lattice and checks whether applying $g$ to the input dataset produces at least one equivalence class of size $k$. 
If so, the algorithm marks $g$ as \emph{anonymous} and computes its information loss using formula~(\ref{formula:loss}), recoding the result for further comparison. 
In case of a tie between the information losses of two distinct states, the algorithm breaks the tie by comparing the \flash criteria. 
Algorithm~\ref{algo:anonymizer} returns the anonymous state with the smallest (optimal) loss.

Algorithm~\ref{algo:flash} uses the anonymizer iteratively to generate a local recoding.
The goal of this algorithm is to retain the maximum possible information of the data by a \emph{greedy} generalization of $D$. 
At each iteration, the algorithm selects a potentially different generalization state, applying it only to the subset of the dataset, forming an equivalence class rather than the entire dataset. 
The chosen generalization state is the one that minimizes information loss among all available options in that iteration.
Any remaining non-anonymous records are retained and processed in subsequent iterations.
In case Algorithm \ref{algo:anonymizer} returns the maximal generalization state \ie $(h^1,\ldots,h^m)$, indicating that no equivalence class can be formed with the remaining records, the loop terminates. 
This process continues until fewer than $k$ records remain in the dataset.
At this point, since these records cannot form a valid equivalence class, they are designated as outliers in the output.
Finally, Algorithm~\ref{algo:flash} returns the locally recoded output, which contains the anonymized equivalence classes and the necessary outliers.

\begin{figure}[H] 
    \flushleft
\includegraphics[scale=0.225]{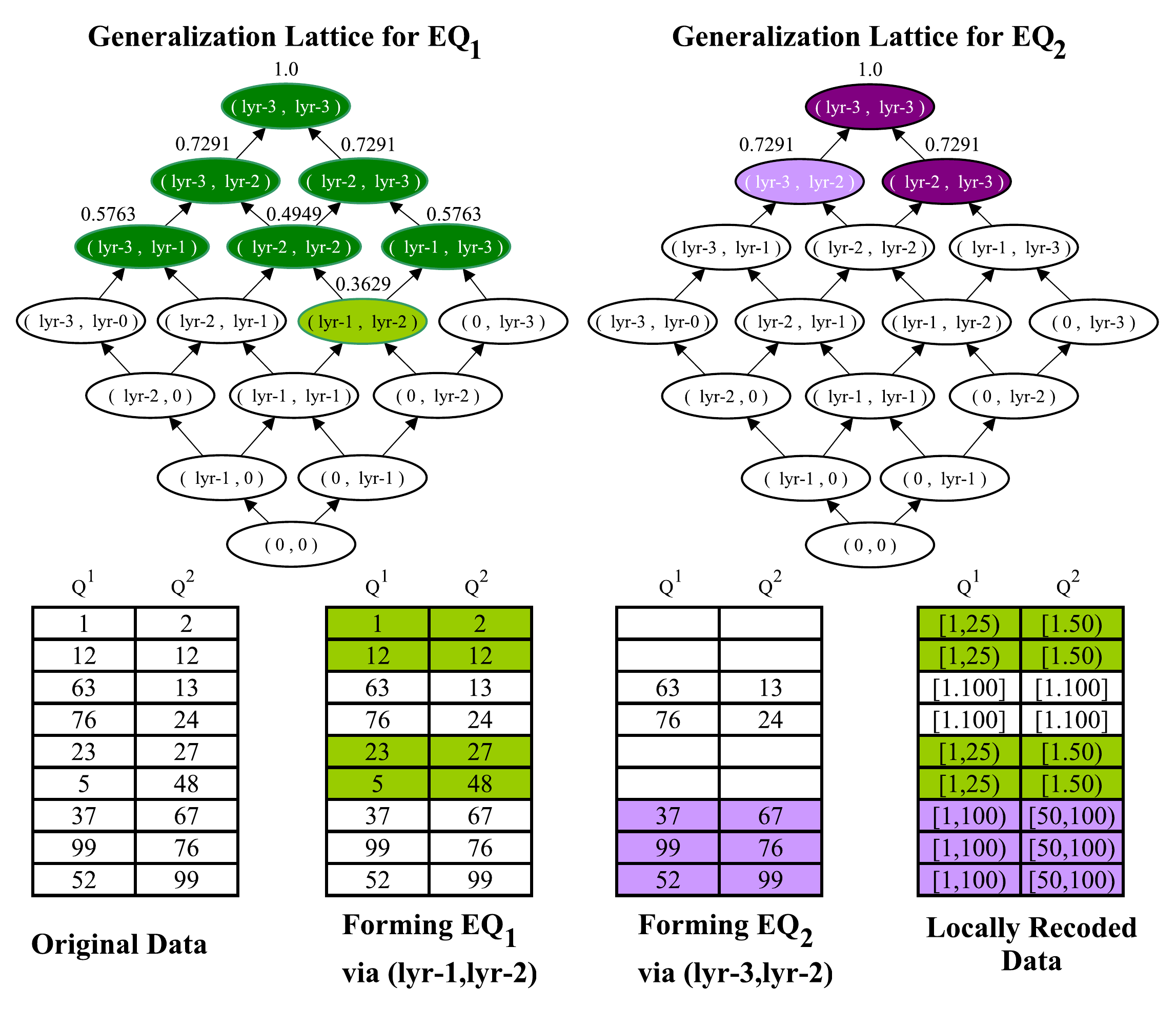}
    \vspace{-0.7cm}
    \caption{From left to right, we show how \flash processes the original dataset to iterate over the generalization lattice, identifying the (colored) states that are anonymous to form equivalence classes $EQ_1$
  and $EQ_2$. Above each anonymous state, the corresponding information loss $loss_g$ is indicated.}
    \label{fig:flash}
\end{figure}

\textbf{An Illustrative Example.} Figure~\ref{fig:flash} illustrates a toy example of applying \flash to a two-dimensional dataset consisting of nine records, with $k = 3$. The original, non-generalized dataset is shown in the leftmost part of the bottom row. 
To construct the first equivalence class, the \flash anonymizer is invoked on the entire original dataset. As it traverses the generalization lattice, the \texttt{\flash-Anonymizer} marks the generalization states that satisfy $k$-anonymity (colored in green) and computes their information loss, displayed above each marked state. Among them, the state $g = (lyr\text{-}1, lyr\text{-}2)$ is selected for having the lowest information loss. Applying this generalization to the original dataset $D$ results in the formation of the equivalence class $EQ_1$, highlighted in green in the bottom row of Figure~\ref{fig:flash}. 
In the next step, the records belonging to $EQ_1$ are removed from $D$, and the remaining non-generalized records are passed to a new invocation of \texttt{\flash-Anonymizer}. 
Given this reduced dataset, the algorithm again traverses the generalization lattice and identifies the anonymous states, this time colored in purple. 
After computing their respective information loss values, a tie is observed between two candidate states. This tie is resolved using the \flash tie-breaking criteria. 
The selected state, $g = (lyr\text{-}3, lyr\text{-}2)$, is then applied to the remaining records, resulting in the formation of $EQ_2$, shown in purple in the bottom row of Figure~\ref{fig:flash}. 
Finally, the two remaining records are insufficient to form an equivalence class of size $k$ and are thus treated as outliers. These outliers are generalized to the root interval of each generalization hierarchy.

\textbf{On the Order of Equivalence Classes.}
At each iteration of the local recoding algorithm, \flash picks the generalization state with the least information loss (breaking ties if needed). 
We emphasize here that there is an \textbf{implicit ordering} on how equivalence classes were formed. 
This means that the equivalence classes generated at the beginning of the local recoding algorithm have lower information loss than those in later iterations.
This ordering information (an intrinsic characteristic of the greedy nature shared by all local recoding algorithms) is one of the key factors contributing to the effectiveness of our proposed CRA attacks.

\section{Combinatorial Refinement Attacks}
In this section, we define the threat model and the objective of the newly proposed combinatorial refinement attack. 
We detail insights on how the greedy choices of \flash lead to inference and, finally, we translate these inferences to linear programming driven CRA.

\subsection{Threat Model \& Definition}

\textbf{Threat Model.} In this threat model, the attacker receives (1) the $k$-anonymous dataset $D_{gen}$ produced by \flash, (2) the anonymity parameter $k$, and  (3) generalization hierarchies $T = (T^1, \ldots, T^m)$.
We emphasize that the attacker has no auxiliary information and neither knowledge nor access to the distribution (or its parameters) used to generate the original dataset $D$.

\textbf{The Definition.} 
Towards defining combinatorial refinement attacks, we will first define the number of quasi-identifier value assignments implied by the original \flash algorithm. 
Recall that a segment (basic or compound) is defined by an interval per quasi-identifier, \ie for quasi-identifier $Q^i$ an associated interval $T^i_{(lyr^i, r^i)}$. 
Suppose, for simplicity, we assume that $Q^i$ can only take integer values, thus, the number of possible quasi-identifier values that $Q^i$ can take are given by the function \texttt{length}$(\cdot)$, \ie, $\texttt{length}\left(T^i_{(lyr^i, r^i)}\right)$. 
More formally, for each segment $S=\Big(T^1_{(lyr^1, r^1)},$ $\ldots,T^m_{(lyr^m, r^m)}\Big)$, associated with an equivalence class $EQ$, the number of value assignments for a single record of $EQ$ is given by:
\begin{displaymath}
\texttt{volume}(S) = \prod_{j=1}^m \texttt{length}\left(T^j_{(l^j, r^j)}\right)
\end{displaymath}

Thus, for a segment $S_{EQ}$ associated with an equivalence class $EQ$ of size $|EQ|$, the number of value assignments for all the records in the class is given by:
\begin{displaymath}
\texttt{LR\_solutions}(EQ) = \binom{\texttt{volume}(S_{EQ})}{|EQ|}.
\end{displaymath}

Informally, a combinatorial refinement attack is successful if it manages to reduce the \texttt{LR\_solutions}$(\cdot)$ metric for at least one of the equivalence classes produced by a local recoding algorithm. 

\begin{definition}
 Let $D_{gen}$ be a $k$-anonymous dataset produced by a local recoding mechanism and $T = (T^1, \ldots, T^m)$ be the corresponding generalization hierarchies. Then, a \emph{combinatorial refinement attack} algorithm $\mathcal{A}$ is successful, if $\mathcal{A}(D_{gen},k,T)$ reduces the number of value assignments for (at least one) equivalence class $EQ$ to be strictly less than $\texttt{LR\_solutions}(EQ)$.
\end{definition}

\textbf{On the Chosen Interpretation of Privacy Guarantees.} A commonly held intuition behind the privacy guarantee of $k$-anonymity is that, given a $k$-anonymized dataset and access to the original data, an adversary cannot re-identify a non-generalized record with probability greater than $1/k$. 
In this work, we examine a different dimension of privacy expectation. 
Specifically, a generalized record (represented as a vector of intervals over quasi-identifiers) is often implicitly understood to mean that any concrete value within each interval is equally plausible.  
This interpretation, while not formally stated in definitions, reasonably reflects how $k$-anonymity is understood by non-experts. 
The goal of the CRA attacker is to demonstrate that not all value assignments within the generalized intervals are truly plausible, given an anonymized dataset.
This exposes a \emph{significant gap} between the perceived privacy guarantees of $k$-anonymity and the actual privacy offered in practice.

\subsection{How Greedy Choices Lead to Inferences}
\label{sec:greedy}

\textbf{Equivalence Classes Cannot Be More ``Fine-Grained''.}
Suppose we run \flash and get an anonymized dataset $D_{\text{gen}}=(y_1,\ldots,y_n)$. 
Additionally, suppose that $D_{\text{gen}}$ contains an equivalence class in which quasi-identifier $Q^i$ has been generalized to the interval $T^i_{(lyr^i,0)}$. 
Notice that since $T^i_{(lyr^i,0)}$ sits in layer $lyr^i$, its interval can be
derived by merging its subintervals $T^i_{(lyr^i-1,0)}$ and $T^i_{(lyr^i-1,1)}$ that reside a layer down, \ie $lyr^i-1$.  
Interestingly, if no non-generalized record falls within one of the subintervals, then $Q^i$ would not be generalized to the parent interval $T^i_{(lyr^i,0)}$. 
This behavior stems from \flash's greedy strategy, which favors intervals that incur a lower information loss. 
Consequently, if an equivalence class can be formed using a more fine-grained node that sits lower in the generalization hierarchy, \flash will prefer that option. 
For example, let $Q^i$ be $T^i_{(2,0)} = [1,50)$.
If the smaller subinterval $T^i_{(1,0)}=[25,50)$ was empty, then $Q^i$ would have been generalized to its sibling $T^i_{(1,1)}=[0,25)$ rather than the larger parent node $T^i_{(2,0)} =[0,50)$. 
The above behavior of \flash implies that both direct subintervals of any interval selected by \flash must contribute toward forming an equivalence class of at least $k$ records. 

Furthermore, we can infer that no single subinterval alone can account for all $k$ records required to form the equivalence class. 
More formally, if either subinterval ($T^i_{(lyr^i-1,0)}$ or $T^i_{(lyr^i-1,1)}$) of the parent interval $T^i_{(lyr^i,0)}$ had contained at least $k$ records, then $Q^i$ would have been generalized to that subinterval instead.
We summarize these key inferences below.

\begin{observation}
Let $EQ$ be an equivalence class formed by the FLASH algorithm where $T^1_{(lyr^1, r^1)},\ldots, T^m_{(lyr^m, r^m)}$ is the list of tree-nodes associated with $EQ$. 
Let $X$ be the non-generalized version of the records from $EQ$. 
Then, for every quasi-identifier  $Q^i$ we have:
\begin{enumerate}[leftmargin=*]
\item for every child-node of $T^i_{(lyr^i, r^i)}$ in layer $lyr^i-1$, there exists at least one $x\in X$ such that $x$ belongs to the interval of this child-node. 
\item for every child-node of $T^i_{(lyr^i, r^i)}$ in layer $lyr^i-1$, there exist at most $k-1$ records of $X$ that belong to the interval of this child node. 
\end{enumerate}
\label{inf:1}
\end{observation}

\textbf{Overlap of Equivalence Classes.}
Suppose we run \flash and get an anonymized dataset $D_{\text{gen}}=(y_1,\ldots,y_n)$. 
Additionally, suppose that $D_{\text{gen}}$ contains an equivalence class $EQ_1$ (that corresponds to a compound segment) and an equivalence $EQ_2$ that can be either a basic or a compound segment. 
It is possible that these segments \emph{overlap} in the $m$-dimensional space. 
Interestingly, the equivalence class (either $EQ_1$ or $EQ_2$) that was formed first during the iterative anonymization of \flash, will ``steal'' the non-generalized records that reside within the overlap. 
This behavior, combined with the implicit ordering of equivalence classes induced by \flash's greedy nature, leads to our next inference: 
Even though two equivalence classes may share one or more segments due to overlap, the data records (if any) within those overlapping segments can only be part of the equivalence class formed first.

\begin{figure}[H] 
    \centering
    \includegraphics[scale=0.18]{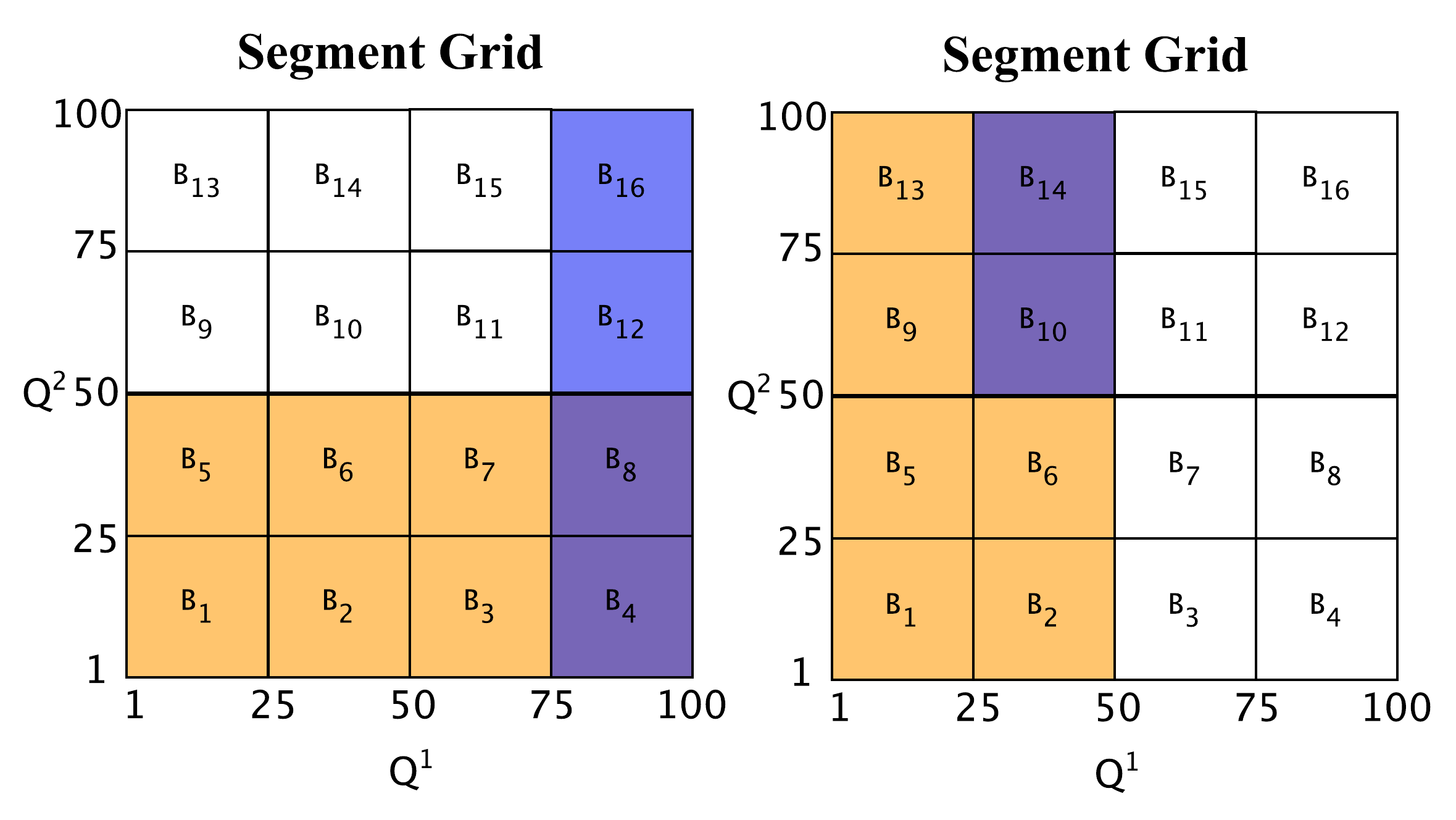}\\
        \vspace{-0.1cm}
      \hspace{0.3cm} (a) \hspace{3cm} (b) 
    \vspace{-0.2cm}
    \caption{(a) The segment grid illustrating two equivalence classes, $EQ_1$ (blue) and $EQ_2$ (orange), with an overlap at basic segments $B_4$ and $B_8$. 
(b) The segment grid illustrating two equivalence classes, $EQ_1$ (blue) and $EQ_2$ (orange), where $EQ_1$ is fully contained within $EQ_2$.}
    \label{fig:attack}
\end{figure}

\begin{observation}
Let $EQ_1$ and $EQ_2$ be two equivalence classes formed by the \flash algorithm by applying generalization states $g_1$ and $g_2$ respectively, such that ($i$) $EQ_1$ was formed before $EQ_2$
and ($ii$) their associated segments are $S_{EQ_1}$ and $S_{EQ_2}$. 
Suppose, there exists a collection of basic segments $B_{\cap}$ that is contained to both $S_{EQ_1}$ and $S_{EQ_2}$. 
Then, if there are any non-generalized records that fall within $B_{\cap}$, they will be generalized as part of $EQ_1$. Consequently, the region associated with $B_{\cap}$ will be empty for $EQ_2$.
\label{inf:2}
\end{observation}
\textbf{Bounds for Non-anonymous Segments.}
Notice that any segment (basic or compound) in the segment grid that does not correspond to a $k$-anonymous equivalence class, must contain strictly fewer than $k$ records. 
If such segments had contained $k$ or more records, they would have formed an equivalence class.

\begin{observation}
Let $D$ be the non-generalized dataset.
Let $\mathcal{E}$ be the set of equivalence classes resulting from running \flash on $D$.
Let $S$ be a segment (basic or compound) that does not correspond to any equivalence class $EQ \in \mathcal{E}$.  
Then, there are at most $k-1$ records from $D$ that reside in segment $S$. 
\label{inf:3}
\end{observation}

\subsection{A New Linear Programming Formulation}

The inferences in Section~\ref{sec:greedy}, drawn purely from the output of \flash, \ie without relying on any assumptions or prior knowledge about the data distribution, offer insights into the location of non-generalized records within the data domain. 
These insights can be formulated as \textbf{bounds on the number of non-generalized records} on segments of the $m$-dimensional data domain. 
Crucially, these bounds emerge as a direct consequence of the decisions made by \flash during local recoding. 
The next step is to rigorously formalize these inferences.

In the following, we leverage these bounds to translate each inference into a \textbf{constraint} within a newly proposed linear programming (LP) formulation for combinatorial refinement attacks. 
In this context, the objective function is not relevant,\ie any feasible solution represents a valid assignment of non-generalized records. 
To formalize the inferences, we introduce a (unknown) variable for each basic segment, representing the number of non-generalized records contained within that basic segment. 
Since basic segments constitute the most fine-grained units of the segment grid, this formulation achieves the highest possible resolution for CRA.

We emphasize that our combinatorial refinement attack constructs \emph{a distinct LP instance for each equivalence class $EQ$ generated by \flash}. The constraints of each LP instance are determined by the relationships of $EQ$ with other segments produced by \flash. 

\textbf{LP Formulation. }Let $\mathcal{B} = \{B_1, \ldots, B_\lambda\}$ be the set of all basic segments in the segment grid.
The total number of basic segments is given by $\lambda = \prod_{i=1}^m 2^{h^i}$,
where $m$ is the number of quasi-identifiers and $h^i$ is the height of the generalization hierarchy $T^i$ for $Q^i$.
To support our formulation, we introduce a vector of counters, denoted in bold as $\mathbf{z} = (z_1, \ldots, z_\lambda)$, where each $z_i$ represents the number of non-generalized records that fall within basic segment $B_i$ while respecting the formed constraints. 
The objective of CRA is to \emph{determine all feasible (integer) assignments for }$\mathbf{z}$ that are all plausible interpretations of the observed anonymization. 
Since each $z_i$ counts the number of records in its corresponding segment $B_i$, it must be a non-negative integer—that is, $z_i \in \mathbb{Z}_{\geq 0}$ for all $i \in {1, \ldots, \lambda}$.

To capture the inferences from Section \ref{sec:greedy} as constraints, we construct a system of inequalities and equalities concerning $\mathbf{z}$:
\begin{displaymath}
\circled{1} \mathbf{A}_{ub}\cdot \mathbf{z}^\top \leq \mathbf{b}_{ub}\text{ , } \circled{2} \mathbf{A}_{lb}\cdot \mathbf{z}^\top \geq \mathbf{b}_{lb} \text{  and  }  \circled{3} \mathbf{A}_{eq}\cdot \mathbf{z}^\top = \mathbf{b}_{eq}.
\end{displaymath}

The left-hand side of each constraint represents a linear combination of the counter variables in $\mathbf{z}$, where the coefficients are specified by the matrices $\mathbf{A}_{ub}$, $\mathbf{A}_{lb}$, and $\mathbf{A}_{eq}$.
Each row in each matrix corresponds to a different constraint.
For a given constraint in the row $j$, we set the coefficient in column $i$ to $1$ if the counter $z_i$ is included in the constraint; otherwise, we set it to $0$.
More formally, $\mathbf{A}_{ub} \in \{0, 1\}^{(p_{ub} \times \lambda)}$,  $\mathbf{A}_{lb} \in \{0, 1\}^{(p_{lb} \times \lambda)}$, and $\mathbf{A}_{eq} \in \{0, 1\}^{(p_{eq} \times \lambda)}$ are binary matrices, where $p_{ub}$, $p_{lb}$, and $p_{eq}$ denote the number of upper bound, lower bound, and equality constraints, respectively. 
For instance, if the $j^{th}$ row of $\mathbf{A}_{ub}$ defines an upper bound over counters corresponding to the compound segment $S$, then $\mathbf{A}_{ub}$ is:
\[
\mathbf{A}_{ub}[j,i] = 
\begin{cases}
1 & \text{, if } B_i \in S\\
0 & \text{, otherwise}
\end{cases}\\
\quad\text{, } \forall j \in [1, p_{ub}], \\ 
i \in [1, \lambda],
\]
The vectors $\mathbf{b}_{ub}$, $\mathbf{b}_{lb}$, and $\mathbf{b}_{eq}$ contain the constant values of the corresponding contraints.
Specifically, the vector $\mathbf{b}_{ub} \in \mathbb{Z}_{\geq0}^{p_{ub}}$ contains the constant for the constraint in \circled{1}, such that  $\mathbf{b}_{ub}[j]$ corresponds to the upper bound of the linear expression $\sum_{i=1}^{\lambda} \mathbf{A}_{ub}[j,i] \cdot z_i$. 
Similarly, the vector $\mathbf{b}_{lb}\in \mathbb{Z}_{\geq0}^{p_{lb}}$ contains the constants of the lower bounds in \circled{2} and the vector $\mathbf{b}_{eq}\in \mathbb{Z}_{\geq0}^{p_{eq}}$ contains the constant of the equality constraint in \circled{3}.

\subsection{CRA on Equivalence Classes}
\label{sec:downcoding_EQ}

\begin{figure}[H] 
    \centering
    \includegraphics[scale=0.18]{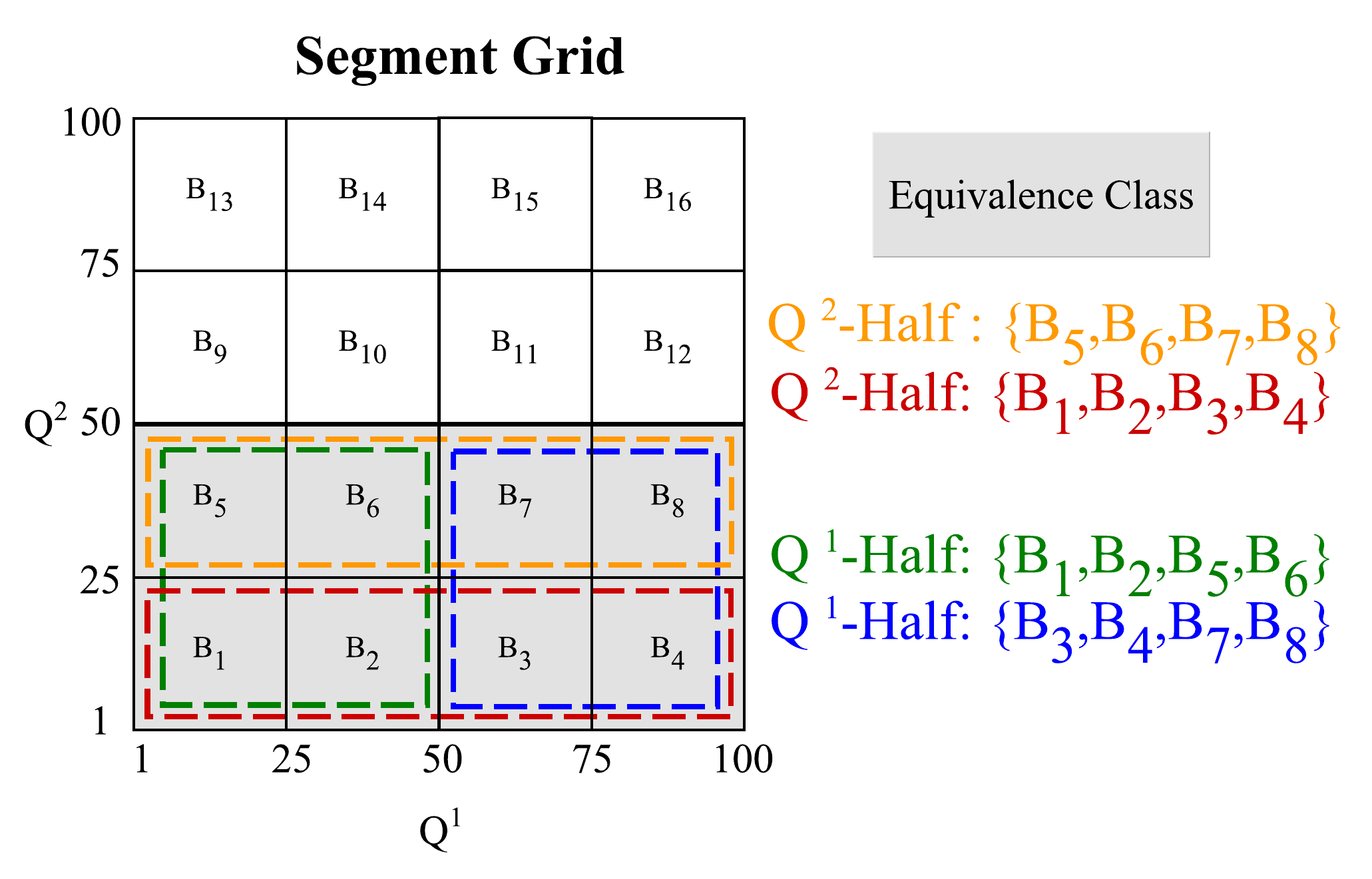}
    \vspace{-0.45cm}
    \caption{Given an equivalence class from \flash (highlighted in gray), the ``Halves Constraint'' requires that each half (two in $Q^1$ dimension and two in $Q^2$) must contain at least one non-generalized data record. Otherwise, \flash would have formed a significantly smaller equivalence class. } 
    \label{fig:halves}
\end{figure}

\textbf{Halves Constraint.} 
Let $EQ$ be a $k$-anonymous equivalence class generated by the \flash algorithm, and let $S$ be the segment associated with it.
According to Inference \ref{inf:1}, for each quasi-identifier $Q^i$ generalized to layer $lyr^i$ in its hierarchy $T^i$, its immediate child nodes at layer $lyr^i -1$ must contain at least one non-generalized record.
This gives rise to what we refer to as \emph{halves constraint}.
To capture this inference, we introduce the notion and generate \emph{"half-segments"} derived from the original compound segment $C$.
Figure~\ref{fig:halves} illustrates an example of half-segments for an equivalence class with generalization state $g = (3, 2)$.
The segment associated with $EQ$ is given by $C = \left(T_{(3,0)}^1, T_{(2,0)}^2\right)$, which can be expressed as the union of the basic segments $C = \bigcup_{t=1}^8 B_t$.
The halves of segment $C$ are shown using dotted rectangles in Figure~\ref{fig:halves} and
can be obtained by ``lowering'' the generalization level of a single quasi-identifier by one layer, while keeping all other layers unchanged. 
When we lower the generalization state of $Q^1$ from $3$ to $2$ we get two half-segments ($i$) $B_1 \cup B_2 \cup B_5 \cup B_6$ and ($ii$) $B_3 \cup B_4 \cup B_7 \cup B_8$.
When we lower the generalization state of $Q^2$ from $2$ to $1$ we get two half-segments ($i$) $B_1 \cup B_2 \cup B_3 \cup B_4$ and ($ii$) $B_5 \cup B_6 \cup B_7 \cup B_8$.

\begin{definition}{[Half-segment]}
\label{def:halves}
Let $EQ$ be an equivalence class for which the generalization state $g = (lyr^1, \ldots, lyr^m)$ was used. Let $C = \left(T_{(lyr^1,r^1)}^1, \ldots,T_{(lyr^m,r^m)}^m\right)$ be the compound segment associated with $EQ$. 
We define $S_h$ to be the \textbf{half-segment} of $C$ constructed by ($1$) choosing an $i\in\{1,\ldots,m\}$ and swapping $T_{(lyr^i,r^i)}^i$ for one of its children in $T^i$ and ($2$) leaving the other tree-nodes of $C$ unchanged. 
\end{definition}
Each quasi-identifier $Q^i$ creates two half-segments, one for each child node in layer $lyr^i-1$.
We set the lower bound for each of the half-segments $S_h$ to at least $1$ non-generalized record.
More formally, the sum of all the basic segments contained in each half-segment $S_h$ must be at least $1$.
To express this in our setting we introduce the $j^{th}$ row of coefficients and the $j^{th}$ element to the constant to the LP instance for $EQ$:
\[
\mathbf{b}_{lb}[j] = 1 \text{ and } 
\mathbf{A}_{lb}[j,i] = \begin{cases}
1 & \text{, if } B_i \in S_h\\
0 & \text{, otherwise}
\end{cases} 
\]
, for all $i\in \{1,\ldots,\lambda\}$. This results in the inequality $\sum_{B_i \in S_h} z_i \geq b_{lb}$.\\

\textbf{Overlap Constraint.} 
Let $EQ_1$ and $EQ_2$ be two equivalence classes generated by the FLASH algorithm such that $EQ_1$ is formed before $EQ_2$.
Let $S_1 \subseteq \mathcal{B}$ and $S_2\subseteq \mathcal{B}$ be the set of basic segments associated with $EQ_1$ and $EQ_2$,  respectively.
If $S_1 \cap S_2 \neq \emptyset$, then $EQ_1$ and $EQ_2$ have an overlap.
In such a scenario, according to Inference \ref{inf:2}, the non-generalized records in the basic segments $S_1 \cap S_2$ must be part of $EQ_1$.
Meanwhile, for $EQ_2$, the basic segments in $S_1 \cap S_2$ must contain no non-generalized records.
To capture this inference, we define an equality constraint for $EQ_2$ that sets the sum of basic segments in the set  $S_1 \cap S_2$ to zero.
More formally, this constraint can be added to the LP formulation for $EQ_2$ by adding row $j$ in the matrix of equality constraints:
\[
\mathbf{b}_{eq}[j] = 0 \text{ and } 
\mathbf{A}_{eq}[j,i] = \begin{cases}
1 & \text{, if } B_i \in S_1 \cap S_2\\
0 & \text{, otherwise}
\end{cases} 
\]
, for all $i\in \{ 1,\ldots,\lambda\}$. This results in the equality $\sum_{B_i \in S_1 \cap S_2} z_i = 0$.\\

To determine the order in which the equivalence classes are formed, we sort them based on their information loss.
In case of a tie, we use the tie-breaking criteria $c_1$ and $c_2$.
We do not use the $c_3$ criterion, as it requires knowledge of the number of distinct values in the generalized and original data $(dst(g(D).Q^i)$ and $dst(D.Q^i))$, which are not available to the attacker under our threat model.
\begin{algorithm}[h!]
        \small
        \caption{\texttt{CRA for Equivalence Classes}\label{algo:downcoding_EQ}}
        \KwData{A $k$-anonymous dataset \texttt{D\_{gen}} produced by \flash, an anonymity parameter $k$, generalization hierarchies $T = (T^1, \ldots, T^m)$}
        \KwResult{All integer solutions \texttt{I} for the LP formulation per equivalence class}
        Initialize an empty set \texttt{I} to store integer solutions across all equivalence class \;
        Extract the set of equivalence classes $\mathcal{E}$ from \texttt{D\_{gen}} \;
        Compute the information loss and criteria ($c_1$ and $c_2$) of each equivalence class in $\mathcal{E}$ using the Equation \ref{formula:loss} and Equation \ref{formula:criteria} respectively\;
        Sort the equivalence classes by ascending information loss, breaking ties using the criteria, and store the result in \texttt{sorted\_EQ}\;
        \ForEach{equivalence class $EQ$ in \texttt{sorted\_EQ} }
        {
            Initialize empty matrices $\texttt{A}_{ub}$, $\texttt{A}_{lb}$, $\texttt{A}_{eq}$ and empty vectors $\texttt{b}_{ub}$, $\texttt{b}_{lb}$ $\texttt{b}_{eq}$\;
            Let $B^*$ be the set of basic segments contained in $EQ$. For each basic segment $B_j$ not in $B^*$, add constraint $z_j=0$ to $\texttt{A}_{eq}, \texttt{b}_{eq}$\;
            \tcp{Overlap Constraints for $EQ$}
            \ForEach{$EQ'$  in \texttt{sorted\_EQ} positioned before $EQ$}
            {
            \If{there are overlapping segments between $EQ'$ and $EQ$}
            {
            Add an overlap constraint to $\texttt{A}_{eq}, \texttt{b}_{eq}$ for segments in $EQ\cap EQ'$\;
            }            
            
            }
            \tcp{Total Sum Constraint for $EQ$}
            Add a total sum constraint to $\texttt{A}_{eq}, \texttt{b}_{eq}$ enforcing $\sum z_i = |EQ|$ \;
            \ForEach{segment $S$ contained in $EQ$}
            {
                \tcp{Halves Constraint for $EQ$}
                \If{$S$ is a half-segment of $EQ$}
                {
                    Add a constraint with bound 1 for $S$ to $\texttt{A}_{lb}, \texttt{b}_{lb}$ \;
                }
            \tcp{Sparse Constraint for $EQ$}
            Add an upper bound constraint for $S$ with bound $k-1$ to $\texttt{A}_{ub}, \texttt{b}_{ub}$ \;            
            }
            Derive all positive integer solutions $\texttt{I}_{EQ}$ for the LP with empty objective function and constraints: $\texttt{A}_{ub} \cdot z \leq \texttt{b}_{ub},\quad
            \texttt{A}_{lb} \cdot z \geq \texttt{b}_{lb},\quad
            \texttt{A}_{eq} \cdot z = \texttt{b}_{eq}$ \tcp*{Solve LP}
            Append $\texttt{I}_{EQ}$ to \texttt{I} \;
        }
        
        Return \texttt{I} \;
\end{algorithm}
\textbf{Sparse Constraints.}
Let $\Phi$ denote the set of all basic and compound segments contained in the equivalence class $EQ$ produced by \flash.
We define a partition of $\Phi$ as (1) the subset $\Phi_{\text{act}}$, which contains all the ``active'' segments, \ie segments that correspond to an equivalence class that is itself contained in $EQ$, and (2) $\Phi_{\neg\text{act}}$ which contains all remaining segments of $\Phi$.
According to Inference \ref{inf:3}, a segment that belongs to the subset $\Phi_{\neg\text{act}}$ must contain strictly less than $k$ non-generalized records, which we call a \emph{sparse constraint}.
More formally, for the LP instance that focuses on $EQ$, we define $\Phi_{\text{act}}$ and $\Phi_{\neg\text{act}}$ so that we generate one constraint for each member $S$ such that $S\in\Phi_{\neg\text{act}}$:
\[
\mathbf{b}_{ub}[j] = k-1  \text{ and }
\mathbf{A}_{ub}[j,i] = \begin{cases}
1 & \text{, if } B_i \in S\\
0 & \text{, otherwise}
\end{cases}
\]
, for all $i\in \{1,\ldots,\lambda\}$. 
This results in $|\Phi_{\neg\text{act}}|$ inequalities of the form $\sum_{B_i \in S} z_i \leq k-1$ which are added to the LP for $EQ$.

\textbf{Total Sum Constraint.} 
A $k$-anonymous equivalence class, namely $EQ$, from the output of the \flash algorithm is constructed by generalizing at least $k$ records to the same generalization state.
This means that the number of generalized records that make up $EQ$ gives us its size.
This gives rise to an equality constraint that we refer to as the \emph{total sum constraint}. 
Let $S_{EQ}$ be the segment that corresponds to $EQ$ in the output of \flash.
Let the number of generalized records in $EQ$ be represented by $|EQ|$.
Then, the sum of all the basic segments that are part of $S_{EQ}$ is equal to $|EQ|$.
We add the following equality constraint to the LP instance for $EQ$:
\[
\mathbf{b}_{eq}[j] = |EQ| \text{ and } 
\mathbf{A}_{eq}[j,i] = \begin{cases}
1 & \text{, if } B_i \in S_{EQ} \\
0 & \text{, otherwise}
\end{cases} 
\]
, for all $i\in \{1,\ldots,\lambda\}$. This results in $\sum_{B_i \in S_{EQ}} z_i = |EQ|$.

\textbf{CRA Algorithm.} 
Algorithm \ref{algo:downcoding_EQ} incorporates all the constraints discussed in Section \ref{sec:downcoding_EQ}.
For each equivalence class $EQ$, we solve a distinct instance of linear programming to identify all the assignments to the counters $\mathbf{z}$ for $EQ$.
Interestingly, a data record can only belong to a single basic segment; therefore, the counters \emph{must only take integer values}. 
Additionally, since we assume no prior auxiliary data about the data distribution, each counter assignment for $\mathbf{z}$ that satisfies the newly discovered constraints is a valid positioning of the non-generalized records. 
Thus, in our attack, we will discover \emph{all integer assignments} for the proposed linear programming problem per equivalence class. 
In theory, integer programming belongs to the NP-complete complexity class~\cite{DBLP:books/fm/GareyJ79}, but in all our tested instances using real data with hundreds of patients, we derived all integer solutions. Notice that for small values of $k$, which are typically preferred in practice, the possible positive integer values are $\{ 0,\ldots,k \}$, significantly limiting the blow-up.

In the Appendix of this work, we present a similar analysis of the LP constraints and the corresponding algorithm for combinatorial refinement attacks on outliers.

\textbf{Breaking $k$-anonymity Definition.}
According to the definition of $k$-anonymity, the requirement is that every record in the anonymized dataset has at least $ k-1$ other records that are indistinguishable across the quasi-identifiers. 
However, the feasible solutions in the CRA output do not satisfy this requirement.
In each CRA output, the records within an equivalence class are assigned to finer-grained ranges (or basic segments in our terminology) instead of sharing the same generalized interval in each dimension.
This results in equivalence classes of size less than $k$, which violates the definition of $k$-anonymity.

\subsection{Quantifying Privacy Reduction from CRA}

Algorithm~\ref{algo:downcoding_EQ} returns a series of data record assignments to segments for \emph{each equivalence class}, \eg focusing on $EQ$, an assignment for basic segments $B_{j}$ and $B_{j+1}$ can be either $(z_1,z_2)=(1,2)$ or $(z_1,z_2)=(2,1)$ for $k=3$, both of which are members of $\texttt{I}_{EQ}$. 
Recall that each segment represents a coarse partitioning of the $m$-dimensional space, so even when a record is assigned to a particular segment, there are multiple possible assignments within it. 
To calculate the number of solutions for $(1,2)\in \texttt{I}_{EQ}$ we have to choose one location for the single record from $B_{j}$ out of the total \texttt{volume}$(B_{j})$ and two locations from $B_{j+1}$ out of the total \texttt{volume}$(B_{j+1})$. 

More formally, for a particular solution $(z_1, \ldots, z_\lambda)$, where $z_i$ records are assigned to segment $B_i$ of volume $\text{volume}(B_i)$, the number of ways to select $z_i$ points from $B_i$ is $\binom{\text{volume}(B_i)}{z_i}$.
Therefore, the number of ways to realize one feasible solution is the product $\prod_{i=1}^\lambda \binom{\text{volume}(B_i)}{z_i}$.
Summing across all feasible solutions $I_{EQ}$ returned for an equivalence class $EQ$ by the CRA algorithm gives the total number of plausible assignments:
\begin{displaymath}
\texttt{CRA\_solutions}(EQ) = \sum_{\mathbf{z} \in I_{EQ}}  \prod_{i=1}^\lambda \binom{\text{volume}(B_i)}{z_i}
\end{displaymath}
We define the \textbf{CRA ratio} as:
\[
\texttt{CRA\_ratio}(EQ) = \frac{\texttt{LR\_solutions}(EQ)}{\texttt{CRA\_solutions}(EQ)}
\]
This ratio captures the relative reduction in uncertainty due to the inferences derived from (1) the greedy decisions of the local recoding algorithm and (2) the observed $k$-anonymous dataset.

\section{Evaluation on Clinical Data}
In this section, we evaluate the effectiveness of the proposed combinatorial refinement attacks on real-world clinical datasets from \clinic, as well as the HCUP dataset. 

\textbf{Datasets.}
For our evaluation, we used two datasets: \texttt{HCUP} (Healthcare Cost and Use Project)~\cite{HCUP} dataset and \clinic dataset.
The selected part of the \texttt{HCUP} dataset contains $1,013$ records and $7$ attributes. 
We selected a subset of four attributes from the dataset for our experiments.
Specifically, the attributes we used are \texttt{GAPICC}, \texttt{APICC}, \texttt{WI\_X}, and \texttt{hosp\_id}, which represent the hospital-specific All-Payer Inpatient Cost-to-Charge Ratio, the group average cost-to-charge ratio, the geographical wage index, and the ID associated with the hospital, respectively.
The \clinic data consists of $500$ records and $6$ attributes that measure various characteristics of the clinical visit. 
These include diastolic blood pressure measurement (\texttt{BP\_Diastolic}), systolic blood pressure measurement (\texttt{BP\_Systolic}), blood oxygen level (\texttt{O2Sat}), body temperature (\texttt{T}), patient weight (\texttt{Wt}), and the unique ID linked to a patient (\texttt{patient\_ID}), which is used in all internal records associated with the patient. 
We chose a subset of 3 attributes for our experiments: \texttt{Wt}, \texttt{BP\_Systolic}, and \texttt{patient\_ID}.
These attributes were chosen because they are commonly present in electronic health records, exhibit sufficient variability for constructing generalization hierarchies, and are plausible quasi-identifiers in clinical datasets.

\textbf{Setup.} 
We anonymized both datasets using the \arx Anonymization Tool, an open-source data anonymization software that implements the \flash algorithm.
Our experiments were conducted using the publicly available codebase of \arx hosted on Github \cite{arx-github}.
\arx supports both local and global recoding algorithms.
However, we selected local recoding in our experiments due to its ability to preserve higher data utility compared to global recoding.
Since all attributes in the datasets are numerical, we defined attribute hierarchies using the \texttt{interval-based hierarchy} setting in \arx.

To enumerate all valid data record assignments for basic segments, we used Google OR-Tools' CP-SAT solver~\cite{google2024ortools}.
Each equivalence class produced by \flash was translated into a linear program with equality and inequality constraints.
We implemented a custom method by extending OR-Tools’ \texttt{CpSolverSolutionCallback} interface. 
Each feasible solution was captured via the OR-Tools method \texttt{on\_solution\_callback()}, which is invoked automatically by the solver during the search process. 
This mechanism, combined with OR-Tools’ \texttt{SearchForAllSolutions()} functionality, enabled exhaustive enumeration of all valid assignments.
To improve scalability, we parallelized the attack using the \texttt{joblib} library, with one solver instance per equivalence class.

Experiments were performed on a computing cluster using the SLURM workload manager.
The experiments were submitted to a compute partition providing access to multiple CPU-cores.
Each job was allocated $4$ CPU cores and $8$ GB of RAM.

\begin{table}[h]
    \centering
    \resizebox{0.48\textwidth}{!}{%
    \begin{tabular}{|c|c|c|c|}
    \hline
    \textbf{Dataset} & \textbf{Dimension} &\textbf{Selected Attributes} & \textbf{Hierarchy Layers} \\
    \hline
    \multirow{3}{*}{\texttt{HCUP}} &2 &hosp\_id, APICC & 3,3\\
    \cline{2-4}
     &3 &hosp\_id, APICC, GAPICC & 3,3,3\\
    \cline{2-4}
     &4 &hosp\_id, APICC, GAPICC, WI\_X & 3,3,3,3\\
    \hline
    \multirow{2}{*}{\clinic} &2 &patient\_weight, BP\_Systolic & 4,3\\
    \cline{2-4}
     &3 &patient\_weight, BP\_Systolic, patient\_ID & 4,3,3\\
    \hline
    \end{tabular}
    }
    \caption{The number of dimensions, the chosen attributes, and the number of layers in each hierarchy for each setup.}
        \label{table:hierarchy_and_dimensions}
\end{table}

\textbf{Methodology.}  
Rather than applying a single CRA instance to the full dataset, we designed multiple experimental configurations to evaluate the robustness and generality of our attack.
We varied the number of quasi-identifiers used in each configuration, with $|QI| \in \{2, 3, 4\}$ for \texttt{HCUP} dataset and $|QI| \in \{2, 3\}$ for \clinic dataset.
For each configuration, we randomly sampled 800 records from the \texttt{HCUP} dataset and 500 records from the \clinic dataset.
In total, we generated 12 independently sampled datasets per configuration to ensure statistical diversity in our evaluation.
For each sampled dataset, we evaluated the performance of CRA for $k$ values ranging from 3 to 7.
The attributes selected for each configuration and the associated number of layers in the hierarchy for each attribute are summarized in Table~\ref{table:hierarchy_and_dimensions}.

\begin{figure}[H] 
    \centering
    \begin{subfigure}[b]{0.435\textwidth}
        \centering
            \includegraphics[width=\linewidth]{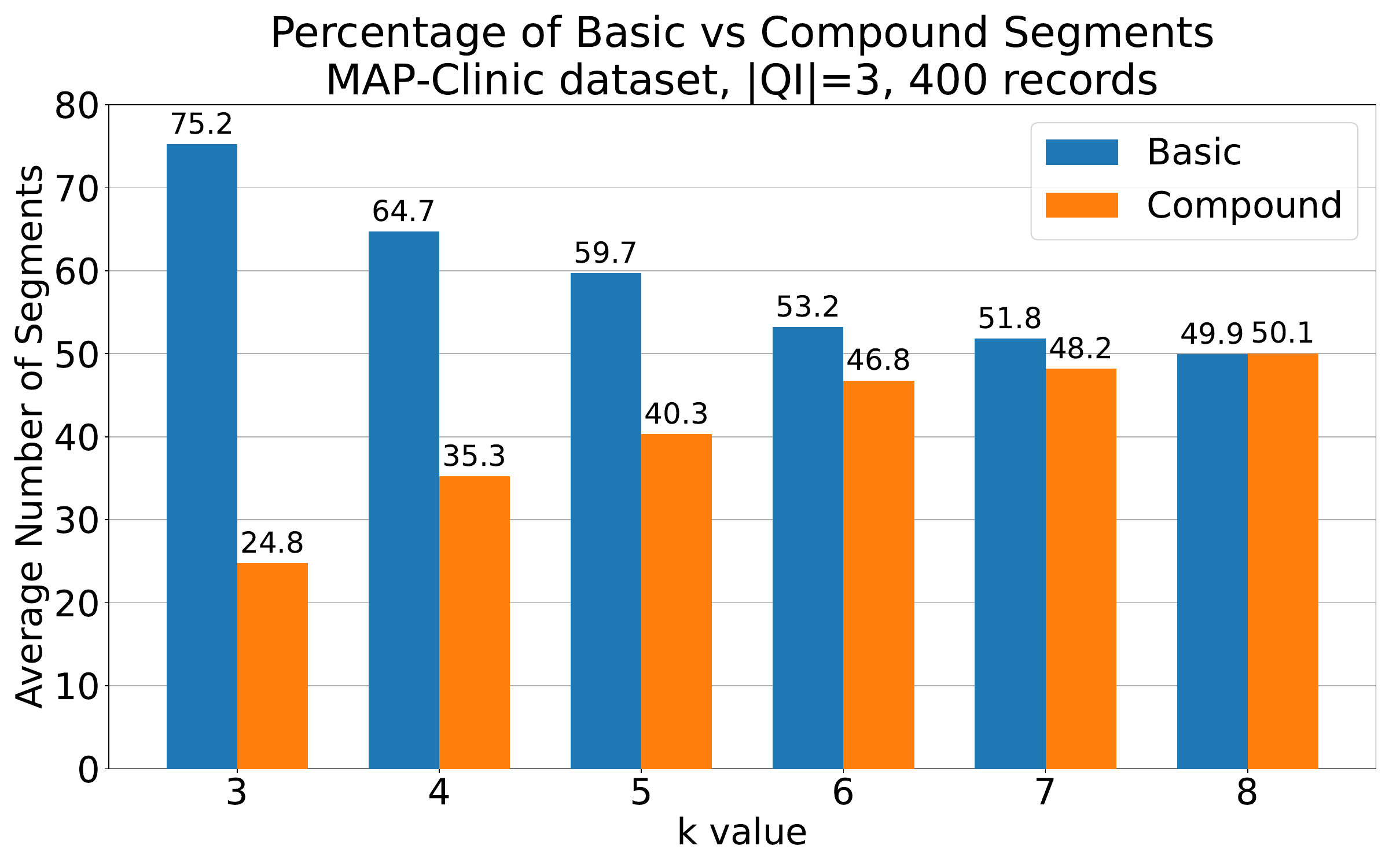}%
        \label{fig:segments1}
    \end{subfigure}

    \begin{subfigure}[b]{0.435\textwidth}
        \centering
            \hspace{-0.3cm}
        \includegraphics[width=\linewidth]{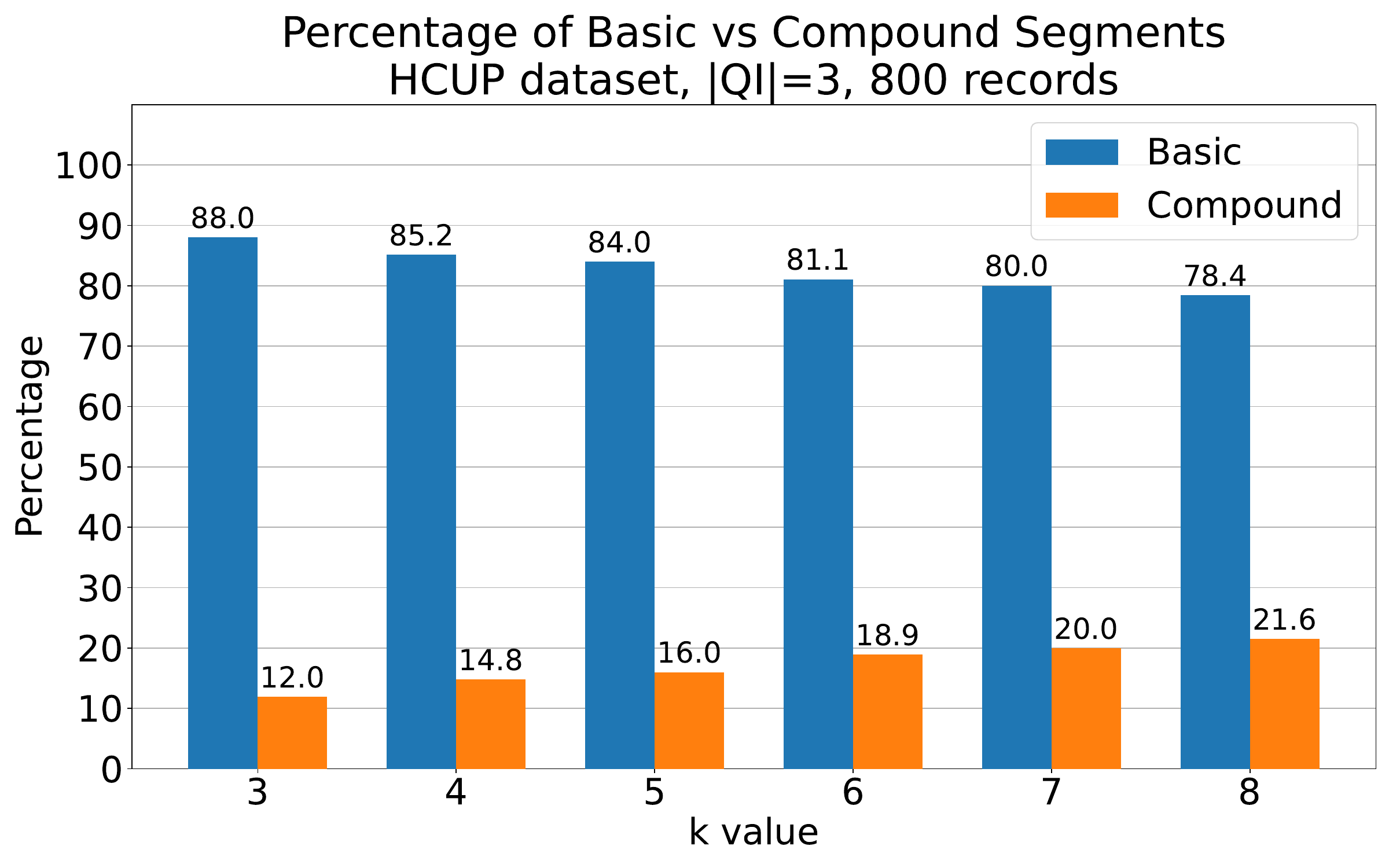}
        \label{fig:segments2}
    \end{subfigure}
    \vspace{-0.2cm}
    \caption{An analysis of the distribution of equivalence classes corresponding to basic segments versus those corresponding to compound segments, for varying values of $k$, on the \clinic and \texttt{HCUP} datasets. }
    \label{fig:segments_percent}
\end{figure}

\begin{table*}[t]
\centering
\renewcommand{\arraystretch}{1.3}
\scriptsize
\resizebox{\textwidth}{!}{%
\begin{tabular}{|c|c||c|c|c||c|c|c||c|c|c||c|c|c||c|c|c||}
\hline
\multirow{2}{*}{\textbf{Dataset}} & \multirow{2}{*}{\textbf{\# QIs}} & \multicolumn{3}{c||}{$k=3$} & \multicolumn{3}{c||}{$k=4$} & \multicolumn{3}{c||}{$k=5$} & \multicolumn{3}{c||}{$k=6$} & \multicolumn{3}{c||}{$k=7$}\\
\cline{3-17}
& & \textbf{LR} & \textbf{CRA} & \colorcell\textbf{CRA} & \textbf{LR} & \textbf{CRA} & \colorcell\textbf{CRA} & \textbf{LR} & \textbf{CRA} & \colorcell\textbf{CRA} & \textbf{LR} & \textbf{CRA} & \colorcell\textbf{CRA} & \textbf{LR} & \textbf{CRA} & \colorcell\textbf{CRA} \\
& & \textbf{Solutions} & \textbf{Solutions} & \colorcell\textbf{Ratio} & \textbf{Solutions} & \textbf{Solutions} & \colorcell\textbf{Ratio} & \textbf{Solutions} & \textbf{Solutions} & \colorcell\textbf{Ratio} & \textbf{Solutions} & \textbf{Solutions} & \colorcell\textbf{Ratio} & \textbf{Solutions} & \textbf{Solutions} & \colorcell\textbf{Ratio}\\
\hline
\hline
\multirow{3}{*}{\texttt{HCUP}} 
& $|QI|=2$ & $8.35{\cdot}10^{15}$ & $8.30{\cdot}10^{14}$ & \colorvalue 6.86 & $2.96{\cdot}10^{22}$ & $1.40{\cdot}10^{21}$ & \colorvalue 6.60 & $4.95{\cdot}10^{29}$ & $8.45{\cdot}10^{27}$ & \colorvalue 7.74 & $1.30{\cdot}10^{30}$ & $5.92{\cdot}10^{28}$ & \colorvalue 7.77 & $2.06{\cdot}10^{36}$ & $4.88{\cdot}10^{34}$ & \colorvalue 11.14 \\
\cline{2-17}
& $|QI|=3$ & $1.43{\cdot}10^{16}$ & $4.37{\cdot}10^{13}$ & \colorvalue 13.34 & $4.05{\cdot}10^{20}$ & $3.96{\cdot}10^{19}$ & \colorvalue 28.70 & $1.76{\cdot}10^{27}$ & $1.24{\cdot}10^{26}$ & \colorvalue 175.31 & $1.74{\cdot}10^{27}$ & $7.37{\cdot}10^{25}$ & \colorvalue 180.60 & $6.68{\cdot}10^{35}$ & $3.22{\cdot}10^{33}$ & \colorvalue 463.07 \\
\cline{2-17}
& $|QI|=4$ &$3.38{\cdot}10^{15}$ &$6.19{\cdot}10^{12}$ & \colorvalue 37.46 &$4.43{\cdot}10^{20}$ &$6.19{\cdot}10^{17}$ & \colorvalue 48.31 &$1.12{\cdot}10^{24}$ &$5.01{\cdot}10^{21}$ & \colorvalue 1,715.76 &$2.71{\cdot}10^{28}$ &$2.16{\cdot}10^{26}$ & \colorvalue 2,700.38 &$6.43{\cdot}10^{30}$ &$5.69{\cdot}10^{27}$ & \colorvalue 10,186.81 \\
\hline
\hline
\multirow{2}{*}{\texttt{\clinic}} 
& $|QI|=2$ & $7.60{\cdot}10^{12}$ & $6.84{\cdot}10^{10}$ & \colorvalue 36.71 & $3.47{\cdot}10^{20}$ & $1.06{\cdot}10^{18}$ & \colorvalue 43.76 & $1.58{\cdot}10^{21}$ & $2.21{\cdot}10^{21}$ & \colorvalue 83.75 & $1.15{\cdot}10^{28}$ & $1.84{\cdot}10^{25}$ & \colorvalue 133.66 & $2.10{\cdot}10^{36}$ & $1.19{\cdot}10^{32}$ & \colorvalue 655.95 \\
\cline{2-17}
& $|QI|=3$ & $5.10{\cdot}10^{22}$ & $3.26{\cdot}10^{21}$ & \colorvalue 13.20 & $1.12{\cdot}10^{39}$ & $1.62{\cdot}10^{36}$ & \colorvalue 33.90 & $1.17{\cdot}10^{45}$ & $7.01{\cdot}10^{41}$ & \colorvalue 50.77 & $1.34{\cdot}10^{56}$ & $1.10{\cdot}10^{51}$ & \colorvalue 797.18 & $1.60{\cdot}10^{62}$ & $1.25{\cdot}10^{60}$ & \colorvalue 1,887.34 \\
\cline{2-17}
\hline
\end{tabular}%
}
\caption{Evaluation of CRA ratio for equivalence classes. Column \texttt{LR} \texttt{solutions} (and \texttt{CRA} \texttt{solutions}) presents the average number of feasible assignments across all equivalence classes in the dataset, averaged over 12 instantiations of the dataset.}
\label{table:equivalence}
\end{table*}
\begin{table*}[t]
\centering
\renewcommand{\arraystretch}{1.3}
\scriptsize
\resizebox{\textwidth}{!}{%
\begin{tabular}{|c|c||c|c|c||c|c|c||c|c|c||c|c|c||c|c|c||}
\hline
\multirow{2}{*}{\textbf{Dataset}} & \multirow{2}{*}{\textbf{\# QIs}} & \multicolumn{3}{c||}{$k=3$} & \multicolumn{3}{c||}{$k=4$} & \multicolumn{3}{c||}{$k=5$} & \multicolumn{3}{c||}{$k=6$} & \multicolumn{3}{c||}{$k=7$}\\
\cline{3-17}
& & \textbf{LR} & \textbf{CRA} & \colorcell\textbf{CRA} & \textbf{LR} & \textbf{CRA} & \colorcell\textbf{CRA} & \textbf{LR} & \textbf{CRA} & \colorcell\textbf{CRA} & \textbf{LR} & \textbf{CRA} & \colorcell\textbf{CRA} & \textbf{LR} & \textbf{CRA} & \colorcell\textbf{CRA}\\
& & \textbf{Solutions} & \textbf{Solutions} & \colorcell\textbf{Ratio} & \textbf{Solutions} & \textbf{Solutions} & \colorcell\textbf{Ratio} & \textbf{Solutions} & \textbf{Solutions} & \colorcell\textbf{Ratio} & \textbf{Solutions} & \textbf{Solutions} & \colorcell\textbf{Ratio} & \textbf{Solutions} & \textbf{Solutions} & \colorcell\textbf{Ratio}\\
\hline
\hline
\multirow{3}{*}{\texttt{HCUP}} 
& $|QI|=2$ &$4.06{\cdot}10^9$ &$1.98{\cdot}10^7$ &\colorvalue 184.05 &$8.84{\cdot}10^{13}$ &$7.81{\cdot}10^{10}$ &\colorvalue 1,377.80 &$2.17{\cdot}10^{18}$ &$4.06{\cdot}10^{14}$ &\colorvalue 9,289.89 &$3.38{\cdot}10^{22}$ &$1.03{\cdot}10^{18}$ & \colorvalue 26,572.58 &$5.87{\cdot}10^{26}$ &$1.78{\cdot}10^{22}$ & \colorvalue 39,049.12 \\
\cline{2-17}
& $|QI|=3$ &$4.64{\cdot}10^{13}$ &$8.80{\cdot}10^{11}$ &\colorvalue 15.75 &$7.79{\cdot}10^{17}$ &$1.16{\cdot}10^{16}$ &\colorvalue 26.78 &$1.68{\cdot}10^{18}$ &$1.31{\cdot}10^{16}$ &\colorvalue 48.17 &$1.84{\cdot}10^{18}$ &$1.15{\cdot}10^{16}$ &\colorvalue 81.68 &$1.62{\cdot}10^{35}$ &$4.57{\cdot}10^{31}$ &\colorvalue 2,470.13\\
\cline{2-17}

& $|QI|=4$ &$8.01{\cdot}10^{13}$ &$2.33{\cdot}10^{12}$ &\colorvalue 8.92 &$1.91{\cdot}10^{18}$ &$2.59{\cdot}t10^{16}$ &\colorvalue 13.78 &$2.10{\cdot}10^{18}$ &$7.41{\cdot}10^{16}$ &\colorvalue 32.14 &$3.36{\cdot}10^{14}$ &$1.50{\cdot}10^{13}$ &\colorvalue 43.41 &$1.20{\cdot}10^{14}$ &$4.19{\cdot}10^{12}$ &\colorvalue 551.11\\
\hline
\hline
\multirow{2}{*}{\texttt{\clinic}} 
& $|QI|=2$ &$4.43{\cdot}10^8$ &$2.93{\cdot}10^7$ &\colorvalue 14.54 &$5.11{\cdot}10^{15}$ &$7.80{\cdot}10^{13}$ &\colorvalue 28.86 &$8.97{\cdot}10^{26}$ &$1.18{\cdot}10^{23}$ &\colorvalue 970.70 &$9.27{\cdot}10^{19}$ &$1.54{\cdot}10^{15}$ &\colorvalue 2,008.49 &$1.12{\cdot}10^{27}$ &$4.67{\cdot}10^{22}$ &\colorvalue 4,274.42 \\
\cline{2-17}
& $|QI|=3$ &$1.53{\cdot}10^{21}$ &$5.98{\cdot}10^{19}$ &\colorvalue 5.40 &$6.97{\cdot}10^{28}$ &$1.65{\cdot}10^{27}$ &\colorvalue 22.76 &$5.93{\cdot}10^{28}$ &$4.46{\cdot}10^{26}$ &\colorvalue 58.47 &$4.05{\cdot}10^{42}$ &$1.07{\cdot}10^{40}$ &\colorvalue 159.25 &$3.29{\cdot}10^{42}$ &$1.63{\cdot}10^{39}$ &\colorvalue 412.18 \\
\hline
\end{tabular}%
}
\caption{Evaluation of CRA ratio for outliers. Column \texttt{LR} \texttt{solutions} (and \texttt{CRA} \texttt{solutions}) presents the average number of feasible assignments across all equivalence classes in the dataset, averaged over 12 instantiations of the dataset.}
\label{table:outliers}
\end{table*}

\subsection{Compound vs. Basic Segments}
In this section, we analyze the distribution of types of segments (basic and compound) associated with equivalence classes in the output of \flash.
Figure \ref{fig:segments_percent} presents this distribution for the \clinic dataset and \texttt{HCUP} datasets.
For this experiment, we report results for up to $k=8$.
For both datasets, we observe a clear trend: as the value of $k$ increases, the proportion of equivalence classes associated with basic segments decreases, while the proportion associated with compound segments increases.
This behavior is consistent with the \flash algorithm, which prioritizes forming equivalence classes with minimal information loss.
To achieve this, \flash favors forming equivalence classes with less generalized intervals, resulting in more basic segments when possible.
For small values of $k$, it is often possible to satisfy the requirement of $k$-anonymity using basic segments. 
Given that the requirement to form an equivalence class is easier to satisfy ($k$ is small), there is a higher chance that at least $k$ records fall within a single basic segment. 
However, as the value of $k$ increases, basic segments may no longer contain enough records to meet this requirement.
As a result, \flash must generalize further by grouping multiple basic segments and forming equivalence classes associated with compound segments.

While both datasets show a decline in the number of basic segments as the value of $k$ increases, the extent of this decline differs between the \clinic dataset and \texttt{HCUP} dataset.
\clinic dataset shows a steep drop of $25.3\%$ from $k = 3$ ($75.2\%$) to $k = 8$ ($49.9\%$).
By $k=8$, the percentage of basic and compound segments for \clinic dataset converge, showing that \flash relies heavily on compound segments to form equivalence classes.
On the other hand, the \texttt{HCUP} dataset shows a much more gradual decline and only decreases by $9.6\%$ over the same range of $k$ values.
We attribute this difference to the dataset size.
The smaller \clinic dataset (400 records) has a lower likelihood of containing $k$ records within a single basic segment, especially as $k$ increases.
Conversely, the larger size of the \texttt{HCUP} dataset (800 records) has more data records that fall within basic segments. 
This enables \flash to form more equivalence classes associated with basic segments.

From the attacker's perspective, the increase in the number of compound segments increases the vulnerability of the anonymized data to Combinatorial Refinement Attacks (CRA).

\begin{table*}[ht]
\centering
\resizebox{\textwidth}{!}{%
\begin{tabular}{|c|c||c|c|c||c|c|c||c|c|c||c|c|c||c|c|c||}
\hline
\multirow{2}{*}{\textbf{Dataset}} & \multirow{2}{*}{\textbf{\# QIs}} & \multicolumn{3}{c||}{$k=3$} & \multicolumn{3}{c||}{$k=4$} & \multicolumn{3}{c||}{$k=5$} & \multicolumn{3}{c||}{$k=6$} & \multicolumn{3}{c||}{$k=7$} \\
\cline{3-17}
& &\textbf{Failed}& \textbf{Single Out} &  \textbf{Single Out} & \textbf{Failed} & \textbf{Single Out} &  \textbf{Single Out} & \textbf{Failed} & \textbf{Single Out} &  \textbf{Single Out} & \textbf{Failed} & \textbf{Single Out} &  \textbf{Single Out} & \textbf{Failed} & \textbf{Single Out} &  \textbf{Single Out}\\

& & \textbf{FPSO} & \textbf{1 Record} &  \textbf{>1 Record} & \textbf{FPSO} & \textbf{1 Record} &  \textbf{>1 Record} & \textbf{FPSO} & \textbf{1 Record} &  \textbf{>1 Record} & \textbf{FPSO} & \textbf{1 Record} &  \textbf{>1 Record} & \textbf{FPSO} & \textbf{1 Record} &  \textbf{>1 Record} \\
\hline
\hline
\multirow{3}{*}{\texttt{HCUP}} 
& $|QI|=2$ & $41.2\%$  & $41.2\%$  & $17.6\%$  & $70.7\%$ & $29.3\%$ & $0\%$  & $82.8\%$ & $15.5\%$ & $1.7\%$ & $72.4\%$ & $24.1\%$ & $3.4\%$ & $74.1\%$ & $19.0\%$ & $6.9\%$\\
\cline{2-17}
& $|QI|=3$ & $12.5\%$ & $62.5\%$ & $25.0\%$  & $29.9\%$ & $39.9\%$ & $30.2\%$  & $51.3\%$ & $22.4\%$ & $26.3\%$ & $48\%$ & $28.2\%$ & $23.8\%$ & $45.1\%$ & $31.7\%$ & $23.2\%$ \\
\cline{2-17}
& $|QI|=4$ & $30.0\%$ & $47.5\%$ & $22.4\%$  & $50.2\%$ & $18.3\%$ & $31.6\%$  & $55.1\%$ & $22.4\%$ & $22.5\%$ & $65.1\%$ & $18.0\%$ & $16.9\%$ & $60.8\%$ & $22.4\%$ &$16.8\%$\\
\hline
\hline
\multirow{2}{*}{\texttt{\clinic}} 
& $|QI|=2$ &$21.6\%$  & $67.2\%$ & $11.2\%$  & $27.2\%$  & $21.9\%$  & $50.9\%$  & $23.8\%$ & $39.7\%$ & $36.5\%$ & $39.4\%$ & $26.3\%$ & $34.3\%$ & $42.0\%$ & $37.0\%$ & $20.9\%$\\
\cline{2-17}
& $|QI|=3$ & $18.3\%$ & $56.6\%$ & $25.1\%$ & $44.1\%$ & $28.7\%$ & $27.2\%$  & $54.6\%$ & $24.6\%$ & $20.7\%$ & $53.4\%$ & $25.2\%$ & $21.4\%$ & $52.9\%$ & $26.2\%$ & $20.9\%$\\
\cline{2-17}
\hline
\end{tabular}%
}
\caption{
The success rate of combining CRA with FPSO for each equivalence class. 
Depending on the outcome of CRA, each equivalence class was categorized as containing $0$, $1$, or $>1$ basic segments with a single record in the CRA-transformed data. 
In all classes with at least one such segment ($1$ or $>1$), the Fuzzy PSO attack always succeeded in singling out a record. The ``\texttt{Failed FPSO}'' column indicates the percentages of equivalence classes where FPSO attack was unsuccessful. 
}
\label{table:pso}
\end{table*}

\subsection{Evaluating CRA Ratio}

\textbf{Higher $k$ Result to Higher Attack Success.} A very interesting phenomenon is observed in
Tables~\ref{table:equivalence} and~\ref{table:outliers}. 
By fixing the number of dimensions/quasi-identifiers for either of the two datasets, we observe that the CRA ratio increases as $k$ increases. 
At first glance, this behavior appears counterintuitive, as an increase in $k$ is typically expected to enhance privacy by grouping more data records together within an equivalence class. 
On the contrary, what we observe is that as $k$ increases, the combinatorial refinement attack becomes more effective, \ie \emph{higher privacy parameter makes the anonymized dataset more vulnerable to privacy attacks}.
This phenomenon can be explained by the fact that higher values of $k$ force the anonymization algorithm to apply more generalization.
As a result, equivalence classes are more likely to be associated with compound segments that span larger portions of the segment grid.
Not only are the segments associated with equivalence classes larger, but they are also more likely to intersect with other equivalence classes to result in overlap.
As a result, the overlap constraint becomes more effective. 
This dramatically decreases the feasible space of original values, leading to a more effective attack.

For equivalence classes (Table~\ref{table:equivalence}), we observe that increasing the number of quasi-identifiers (\ie moving to higher-dimensional data) further amplifies the effectiveness of CRA. 
A potential reason for this could be the increased number of half-segments introduced with each additional quasi-identifier. 
According to Definition~\ref{def:halves}, each quasi-identifier contributes two half segments. 
Therefore, as the dimensionality of the data increases, the number of half segments grows linearly with the number of quasi-identifiers. 
This results in more \emph{half-constraints} being added to the CRA formulation, which in turn further restricts the feasible region of solutions.
These tighter constraints lead to better refinement of the non-generalized records, leading to a more effective attack.

Interestingly, the outliers do not present the same trends as the equivalence classes. 
As shown in Table~\ref{table:outliers}, increasing the number of quasi-identifiers does not consistently lead to higher CRA effectiveness for outliers. 
This phenomenon appears because as the number of dimensions increases, the number of basic segments that participate in an equivalence class decreases. 
As a result, the overlap constraints in higher dimensions generate more feasible solutions (due to a smaller number of constraints), which leads to a drop in the CRA ratio. 
Recall that the number of basic segments grows exponentially with the number of dimensions. 
Thus, in the \texttt{HCUP} data for $k=6$, the average overlap drops from $88.1\%$ ($\sim$14 out of 16 basic segments) for two dimensions to $77.0\%$ ($\sim$49 out of 64 basic segments) for three dimensions.
Another potential cause for this decrease could be the lack of half-constraints.
As discussed in Appendix~\ref{appendix:1}, if the number of outliers is fewer than $k$, then the halves constraint is not applicable.
The absence of these constraints means that the restrictions on the feasible region do not increase with increasing dimensions, like in the case of equivalence classes.
\textbf{Number of CRA Assignments.}
Table \ref{table_combinations} reports the average number of CRA output assignments per equivalence class across values of $k$ and quasi-identifier dimensions.
The table reveals the effectiveness of the attack: although millions of raw data value assignments are theoretically possible under $k$-anonymity, CRA prunes this space down to a small set of plausible assignments. 
For instance, at $k = 3$ and $|QI| = 4$, CRA reduces (on average) the space to just $593.86$ plausible integer assignments (HCUP). 
Even when the number of assignments is large (e.g., $274,874$ at $|QI|=3$, $k=7$ for \clinic), it is still drastically smaller than the full product of the range in each dimension of the equivalence class.

\begin{table}[t]
\centering
\scriptsize
\resizebox{0.48\textwidth}{!}{%
\begin{tabular}{|c|c||c|c|c|c|c|}
\hline
\multirow{2}{*}{\textbf{Dataset}} & \multirow{2}{*}{\textbf{\# QIs}} & \multicolumn{5}{c|}{\textbf{CRA Combinations}} \\
\cline{3-7}
& & $k=3$ & $k=4$ & $k=5$ & $k=6$ & $k=7$ \\
\hline
\hline
\multirow{3}{*}{\texttt{HCUP}} 
& $|QI|=2$ & $1.72$ & $2.33$ & $4.91$ & $10.08$ & $16.36$ \\
\cline{2-7}
& $|QI|=3$ & $21.98$ & $53.89$ & $543.85$ & $1,968.66$ & $31,229.73$ \\
\cline{2-7}
& $|QI|=4$ & $593.86$ & $24,461.10$ & $145,945.35$ & $1,032,824.70$ & $895,620.66$ \\
\hline
\hline
\multirow{2}{*}{\texttt{\clinic}} 
& $|QI|=2$ & $3.51$ & $26.42$ & $89.83$ & $298.63$ & $941.45$ \\
\cline{2-7}
& $|QI|=3$ & $36.46$ & $1,032.60$ & $17,138.35$ & $33,354.02$ & $274,874.76$ \\
\hline
\end{tabular}%
}
\caption{Number of CRA combinations for equivalence classes. Each column presents the average number of combinations across all equivalence classes in the dataset, averaged over 12 instantiations of the dataset.}
\label{table_combinations}
\end{table}

\begin{table}[t]
\centering
\scriptsize
\resizebox{0.48\textwidth}{!}{%
\begin{tabular}{|c|c||c|c|c|c|c|}
\hline
\multirow{2}{*}{\textbf{Dataset}} & \multirow{2}{*}{\textbf{\# QIs}} & \multicolumn{5}{c|}{\textbf{Average Runtime (in seconds)}} \\
\cline{3-7}
& & $k=3$ & $k=4$ & $k=5$ & $k=6$ & $k=7$ \\
\hline
\hline
\multirow{3}{*}{\texttt{HCUP}} 
& $|QI|=2$ & $8.0\cdot 10^{-4}$ & $6.6\cdot 10^{-4}$ & $7.9\cdot 10^{-4}$ & $9.9\cdot 10^{-4}$ & $1.2\cdot 10^{-3}$ \\
\cline{2-7}
& $|QI|=3$ & $7.8\cdot 10^{-3}$ & $9.9\cdot 10^{-3}$ & $6.2\cdot 10^{-2}$ & $8.5\cdot 10^{-1}$ & $1.4\cdot 10^{2}$ \\
\cline{2-7}
& $|QI|=4$ & $4.5\cdot 10^{-1}$ & $1.1\cdot 10^{2}$ & $3.3\cdot 10^{2}$ & $3.9\cdot 10^{3}$ & $4.1\cdot 10^{3}$ \\
\hline
\hline
\multirow{2}{*}{\texttt{\clinic}} 
& $|QI|=2$ & $1.4\cdot 10^{-3}$ & $3.3\cdot 10^{-3}$ & $7.3\cdot 10^{-3}$ & $2.3\cdot 10^{-2}$ & $8.7\cdot 10^{-2}$ \\
\cline{2-7}
& $|QI|=3$ & $1.0\cdot 10^{-2}$ & $2.4\cdot 10^{-1}$ & $6.1$ & $3.3\cdot 10^{1}$ & $3.4\cdot 10^{1}$ \\
\hline
\end{tabular}%
}
\caption{Average CRA Runtime per equivalence class}
\label{table_runtime}
\end{table}

\textbf{CRA Runtime.}

Table \ref{table_runtime} reports the average runtime (in seconds) for solving an instance of CRA linear programming for a single equivalence class.
The runtime increases with both $k$ and the number of quasi-identifiers.
As $k$ increases, we observe that \flash uses more basic segments in order to identify at least $k$ records to form an equivalence class. 
In turn, the increase in basic segments results in an increase in the number of variables associated with the linear programming formulation.
Consequently, solving the linear program becomes more computationally expensive.
Analogously, increasing the number of quasi-identifiers results in a higher number of dimensions, which also increases the number of variables, leading to longer run times. 
In the Appendix \ref{section:outlier_runtime} of this work, we present the runtime for outliers in Table \ref{table_runtime_outlier}.

\section{Future Direction: From CRA to ``Fuzzy'' PSO}

Even though CRA is an attack on the privacy of $k$-anonymity (since the $k$-anonymity definition is violated), in this section, we explore how CRA can serve as a component of a different attack. 
We emphasize that this extension is not the focus of our work and is presented as a direction for future work. 
The threat model in this section is similar to CRA in that it also has no access to auxiliary information. 

Specifically, the multi-stage attack that we propose applies CRA in the first stage and, based on the returned assignments, forms a \emph{Fuzzy} Predicate Singling Out (or FPSO) attack based on the original PSO proposed in~\cite{Altman2021Hybrid, DBLP:journals/pnas/CohenN20}. 
The objective of the Fuzzy PSO attack is to uniquely identify a \emph{single} individual in the non-anonymized dataset. 
It does so by using the anonymized dataset to form a \emph{set} of predicates, as opposed to traditional PSO that identifies a single predicate. 
Out of this set of predicates, only one of them successfully singles out a record in the non-anonymized dataset.  
In other words, this predicate evaluates true only for a single record in the non-anonymized dataset $D$. 

\textbf{Only One of the CRA Outputs is Valid.}
We say that a CRA output assignment $\mathbf{z} = (z_1, \ldots, z_\lambda)$ is considered \emph{valid} with respect to non-anonymized dataset $D$ if~\footnote{When $z_i=0$, we have two possibilities, either (1) there are no records in the non-anonymized dataset that fall within basic segment $B_i$ or (2) there are records in basic segment $B_i$ but they have been ``stolen'' by an earlier-formed equivalence class.} for every $z_i>0$, there exist exactly $z_i$ records in the non-anonymized dataset that fall within basic segment $B_i$. 
If two or more CRA output assignments were valid, this would imply that two disjoint sets of records formed two distinct equivalence classes, in different iterations of \flash, both of which correspond to the same compound segment. 
This contradicts the way local recoding is performed in \flash, which forms an equivalence class by including \emph{all records} that fall within the chosen compound segment.
The above argument shows that only one assignment among the CRA outputs can be valid.

\textbf{Forming a FPSO from CRA.} 
Following this insight, we test every output of CRA against the non-anonymized dataset $D$ so as to confirm that a valid output exists (much like the PSO attack, which applies its predicate to the non-anonymized dataset).

This process eliminates all invalid assignments of records to basic segments, leaving us with only a single valid assignment of records to basic segments. 
We note that the described approach is not a typical PSO approach~\cite{Altman2021Hybrid, DBLP:journals/pnas/CohenN20} in which a single predicate is identified first and then applied on $D$ for verification. 
Instead, here we use multiple ``candidate'' predicates based on CRA's outputs, but with the knowledge that only one of them can act as a traditional PSO predicate.

Once the valid assignment $\mathbf{z} = (z_1, \ldots, z_\lambda)$ is located, we examine its basic segment counters. 
For all $z_i=1$, we can infer that the corresponding basic segment $B_i$ contains exactly one record in the non-anonymized dataset. 
This means the segment ranges across quasi-identifiers, uniquely isolating that individual within the full dataset $D$.
More formally, a basic segment
\begin{displaymath}
B = \left(T^1_{(1, r^1)},\ldots,T^m_{(1, r^m)}\right)\text{,  where } r^i \in [0,2^{(h^i-1)}-1]
\end{displaymath}
isolates a data record $x = (x^{1},\ldots,x^{m})$ if the predicate $P$ defined as
\begin{displaymath}
\left( (x^{1} \in T^1_{(1, r^1)}) \land \ldots \land (x^{m} \in T^m_{(1, r^m)})\right)\text{,  where } r^i \in [0,2^{(h^i-1)}-1]
\end{displaymath}
evaluates to true for only one record in  $D$.
This is consistent with what the Article 29 Working Party~\cite{Article29Opinion2014} refers to as ``\emph{narrowing down [to a singleton] the group to which [the individual] belongs}'' by specifying ``\emph{criteria which allows him to be recognized}'' an argument also made in~\cite{DBLP:journals/pnas/CohenN20}. 
In case $z_i\neq 1$ for all $i$, then the Fuzzy PSO attack fails since there are multiple records within the corresponding $B_i$ and we cannot deterministically single out any of them. 

\textbf{Experiments.} To quantify the effectiveness of extending CRA to Fuzzy PSO attacks, we conducted experiments on the HCUP and \clinic datasets. 
As a sanity check, we experimentally confirmed that among all CRA output assignments, only a single assignment was valid (with respect to the corresponding $D$). 
Interestingly, given a valid assignment $\mathbf{z} = (z_1, \ldots, z_\lambda)$, there can be multiple counters with value $1$. 
We consider a Fuzzy PSO attack successful if there exists at least one counter with value $1$.

Given that the aforementioned property is highly data-dependent, we recorded the percentage of equivalence classes that had ($i$) no counters with value $1$ which means an unsuccessful FPSO, ($ii$) exactly one counter with value $1$ which means that exactly one record was isolated, and ($iii$) more than one counters with value $1$ which means multiple records were singled out.
The results are reported in Table~\ref{table:pso}. 
In our experiment, we observed varying degrees of success where some parameterizations resulted in $\sim20\%$ of equivalence classes being susceptible to FPSO and other parameterizations resulted in $\sim90\%$ of equivalence classes being vulnerable to Fuzzy PSO. 
These results demonstrate that CRA can enable Fuzzy PSO attacks, significantly increasing the adversary’s ability to probabilistically isolate individuals even under strong anonymization parameters.
We leave for future work the investigation of whether the FPSO attack can be strengthened with distributional knowledge.

\section{Limitations}
While Combinatorial Refinement Attack highlights major privacy risks in the local recoding algorithm of \arx, it is important to acknowledge its limitations. 
First, it is unclear whether the proposed CRA is applicable to local recoding anonymization algorithms that use randomized or non-greedy strategies. 
All the proposed inferences in this work are based on the greedy nature of local recoding. 
Second, CRA is not applicable to global recoding algorithms, where the same generalization is applied uniformly across the entire dataset. 
Third, the scalability of CRA depends on the number of quasi-identifiers and the depth of the generalization hierarchy.
As the number of quasi-identifiers and/or the depth of the hierarchy increases, the number of variables in the underlying linear programming formulation grows.
Finally, it is unclear if the proposed CRA can be applied as is to other local recoding packages (e.g., sdcMicro, Amnesia).
We hypothesize that as long as their local recoding approach is greedy, similar ideas to the ones presented here can be applied in these packages.

\section{Conclusion}
In this work, we introduce a new family of attacks that challenge the privacy expectations commonly associated with local recoding of $k$-anonymity. 
In generalized datasets, numerical attributes are commonly replaced with intervals intended to represent a range of plausible values. 
Our Combinatorial Refinement Attacks reveal that many of these values are, in practice, not plausible. 
Our findings highlight a significant mismatch between the privacy that users expect and the protection actually offered by locally recoded $k$-anonymized data. 
Notably, our techniques require no auxiliary information, a key distinction from all prior attacks in this field. 

Overall, our findings highlight that even decades-old privacy techniques (such as $k$-anonymity, introduced nearly 30 years ago) still admit rigorous audit and continue to reveal previously overlooked privacy vulnerabilities.
\section*{Acknowledgements}

The project was supported by the Commonwealth Cyber Initiative (CCI) grant from the program \emph{``Securing Interactions between Humans and Machines''}. Partial support for the first and fourth authors was provided by NSF Award \#2154732. The authors thank Tessa Joseph for valuable discussions during the early stages of this work and Anthony Wiest for assistance with data extraction. 
The authors thank the \texttt{ARX} team for their constructive feedback during the disclosure process and for their valuable technical insights.

{\footnotesize \bibliographystyle{acm}
\bibliography{tex/references}}
\appendix 
\clearpage
\section{Appendix}
\subsection{CRA on Outliers}
\label{appendix:1}
Outliers can be thought of as a special case of equivalence classes that have been generalized to the highest generalization state $g = (h^1,\ldots,h^m)$. 
The segment associated with outliers is the entire segment grid.
Unlike $k$-anonymous equivalence classes chosen by \flash, there can exist fewer than $k$ records suppressed to form outliers. 
However, most of the constraints that apply to equivalence classes also apply to outliers, with slight modifications.

\textbf{Halves Constraint.}
Let $O$ represent the set of outliers. When the number of outliers is greater than or equal to $k$, \ie $|O| \geq k$, we can apply the halves constraint.
This is because, if one half of the segment grid had zero records, the other half would contain all $|O| \geq k$ records. 
In that case, the algorithm would have formed an equivalence class using that half-segment instead of suppressing those records to form outliers. 
Let $S_H$ be a half-segment constructed by generalizing one quasi-identifier $Q^i$ to the layer $h^i - 1$, while all other quasi-identifiers remain generalized to the top layer $h^j$. 
For each such half-segment $S_H$, we require that it contains at least one non-generalized record from $O$.
This constraint can be added to the LP formulation for $O$ by adding row $j$ to coefficients and $j^{th}$ constant as follows:
\[
\mathbf{b}_{lb}[j] = 1 \text{ and } 
\mathbf{A}_{lb}[j,i] = \begin{cases}
1 & \text{, if } B_i \in S_H\\
0 & \text{, otherwise}
\end{cases} 
\]
, for all $i\in \{ 1,\ldots,\lambda\}$. This results in the equality $\sum_{B_i \in S_H} z_i \geq 1$.

\textbf{Overlap Constraint.} 
Since outliers have the maximum possible information loss, they are formed last -- only after all the equivalence classes have been generated.
As a result, any equivalence class $EQ$ ``steals"  all non-generalized records that lie within the segment representing $EQ$.
More formally, let $G$ represent the segment grid.
Let $\mathcal{E}$ represent the set of all equivalence classes.
If $G \cap \mathcal{E} \neq \emptyset$, then by Inference~\ref{inf:2}, the area of the overlap should not contain any outliers.
This constraint can be added to the LP formulation for $O$ by adding row $j$ to coefficients and $j^{th}$ constant as follows:
\[
\mathbf{b}_{eq}[j] = 0 \text{ and } 
\mathbf{A}_{eq}[j,i] = \begin{cases}
1 & \text{, if } B_i \in G \cap \mathcal{E}\\
0 & \text{, otherwise}
\end{cases} 
\]
, for all $i\in \{ 1,\ldots,\lambda\}$. This results in the equality $\sum_{B_i \in G \cap \mathcal{E}} z_i = 0$.

\textbf{Bounding Constraints.} 
Bounding constraint for outliers follow the same principle as that for equivalence classes, with a slight modification. 
Unlike equivalence classes, outliers are generalized to span the entire segment grid.
As a result, in the case of outliers, $\phi$ denotes the set of all basic and compound segments in the segment grid, $\phi_{act}$ contains all the segments that represent an equivalence class in \flash output, and $\phi_{\neg act}$ contains the remaining segments from the segment grid.
According to inference \ref{inf:3}, a segment that belongs to the subset $\phi_{\neg act}$ must contain less than $k$ non-generalized records. 
Therefore, we generate a constraint for every $S$ such that $S \in \phi_{\neg act}$:
\[
\mathbf{b}_{ub}[j] = k-1  \text{ and }
\mathbf{A}_{ub}[j,i] = \begin{cases}
1 & \text{, if } B_i \in S\\
0 & \text{, otherwise}
\end{cases}
\]
, for all $i\in \{1,\ldots,\lambda\}$. 
This results in $|\Phi_{\neg\text{act}}|$ inequalities of the form $\sum_{B_i \in S} z_i \leq k-1$ which are added to the LP for outliers.

\textbf{Total Sum Constraint.}
This constraint applies to outliers just as it does to equivalence classes.
Let $O$ represent the set of outliers such that the number of outliers is given by $|O|$.
Let the segment grid be represented by $G$.
The sum of all basic segments in the segment grid should be equal to $|O|$.
We add the following equality constraint to the LP instance that attacks outliers:
\[
\mathbf{b}_{eq}[j] = |O| \text{ and } 
\mathbf{A}_{eq}[j,i] = \begin{cases}
1 & \text{, if } B_i \in G \\
0 & \text{, otherwise}
\end{cases} 
\]
, for all $i\in \{1,\ldots,\lambda\}$. This gives rise to the equality $\sum_{B_i \in G} z_i = |0|$.

\textbf{Algorithm.} 
Algorithm \ref{algo:downcoding_outliers}, incorporates all the constraints discussed in Appendix \ref{appendix:1}.
We use linear programming to identify all the assignments to the counters $\mathbf{z}$ for outliers.

\begin{algorithm}[h!]
        \small
        \caption{\texttt{CRA for Outliers}\label{algo:downcoding_outliers}}
        \KwData{A $k$-anonymous dataset \texttt{D\_{gen}} produced by \flash, an anonymity parameter $k$, generalization hierarchies $T = (T^1, \ldots, T^m)$}
        \KwResult{All integer solutions for outliers $\texttt{I}_{out}$}
        Initialize an empty set $\texttt{I}_{out}$ to store all integer solutions for outliers\;
        Extract the set of equivalence classes $\mathcal{E}$ from \texttt{D\_{gen}} \;
        Initialize empty matrices $\texttt{A}_{ub}$, $\texttt{A}_{lb}$, $\texttt{A}_{eq}$ and empty vectors $\texttt{b}_{ub}$, $\texttt{b}_{lb}$ $\texttt{b}_{eq}$\;
        Get the segment grid $G$ using the generalization hierarchies.\;
        \tcp{Overlap Constraints for outliers}
        \ForEach{EQ in $\mathcal{E}$}
        {
            Add an overlap constraint to $\texttt{A}_{eq}, \texttt{b}_{eq}$ for segments in $EQ\cap G$\;
        }
        \tcp{Total Sum Constraints for outliers}
        Add a total sum constraint to $\texttt{A}_{eq}, \texttt{b}_{eq}$ enforcing $\sum_{z_i \in G} z_i = \text{number of outliers}$ \;
        \ForEach{segment $S$ contained in $G$}
        {
            \tcp{Halves Constraints for outliers}
            \If{$S$ is a half-segment of $EQ$ and number of outliers $\geq k$}
            {
                Add a constraint with bound 1 for $S$ to $\texttt{A}_{lb}, \texttt{b}_{lb}$ \;
            }
            \tcp{Sparse Constraint for outliers}
            \If{$S\in \phi_{\neg act}$}
            {
                Add an upper bound constraint for $S$ with bound $k-1$ to $\texttt{A}_{ub}, \texttt{b}_{ub}$ \;
            }
        }
        Derive all positive integer solutions $\texttt{I}_{out}$ for the LP with empty objective function and constraints: $\texttt{A}_{ub} \cdot z \leq \texttt{b}_{ub},\quad
            \texttt{A}_{lb} \cdot z \geq \texttt{b}_{lb},\quad
            \texttt{A}_{eq} \cdot z = \texttt{b}_{eq}$ \tcp*{Solve LP}
        Return $\texttt{I}_{out}$\;
\end{algorithm}

\subsection{Experiments}

\begin{figure}[H] 
    \centering
    \begin{subfigure}[b]{0.45\textwidth}
        \centering
        \includegraphics[width=\linewidth]{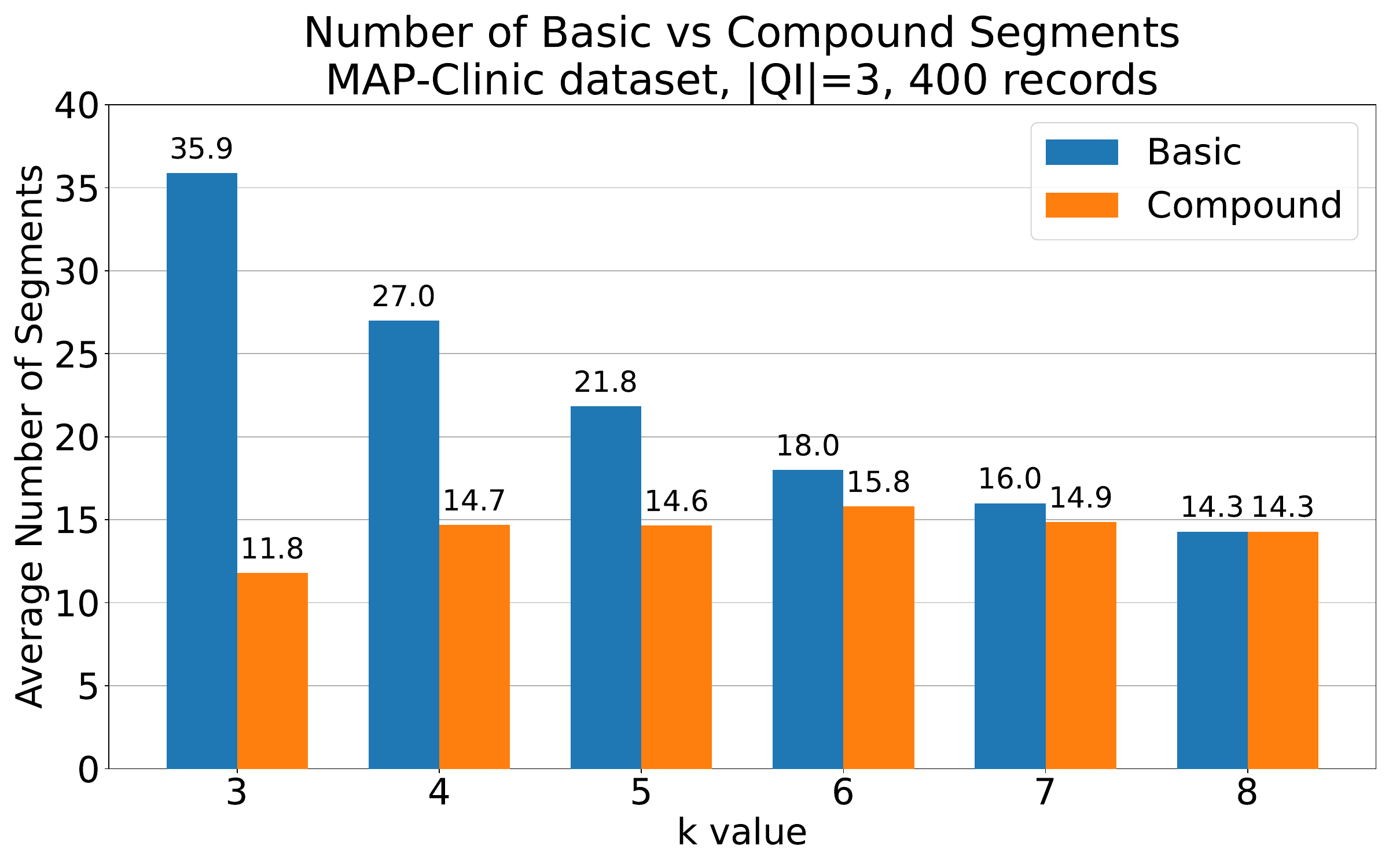}
        \label{fig:segments1_count}
    \end{subfigure}
    \vfill
    \begin{subfigure}[b]{0.45\textwidth}
        \centering
            \hspace{-0.3cm}
        \includegraphics[width=\linewidth]{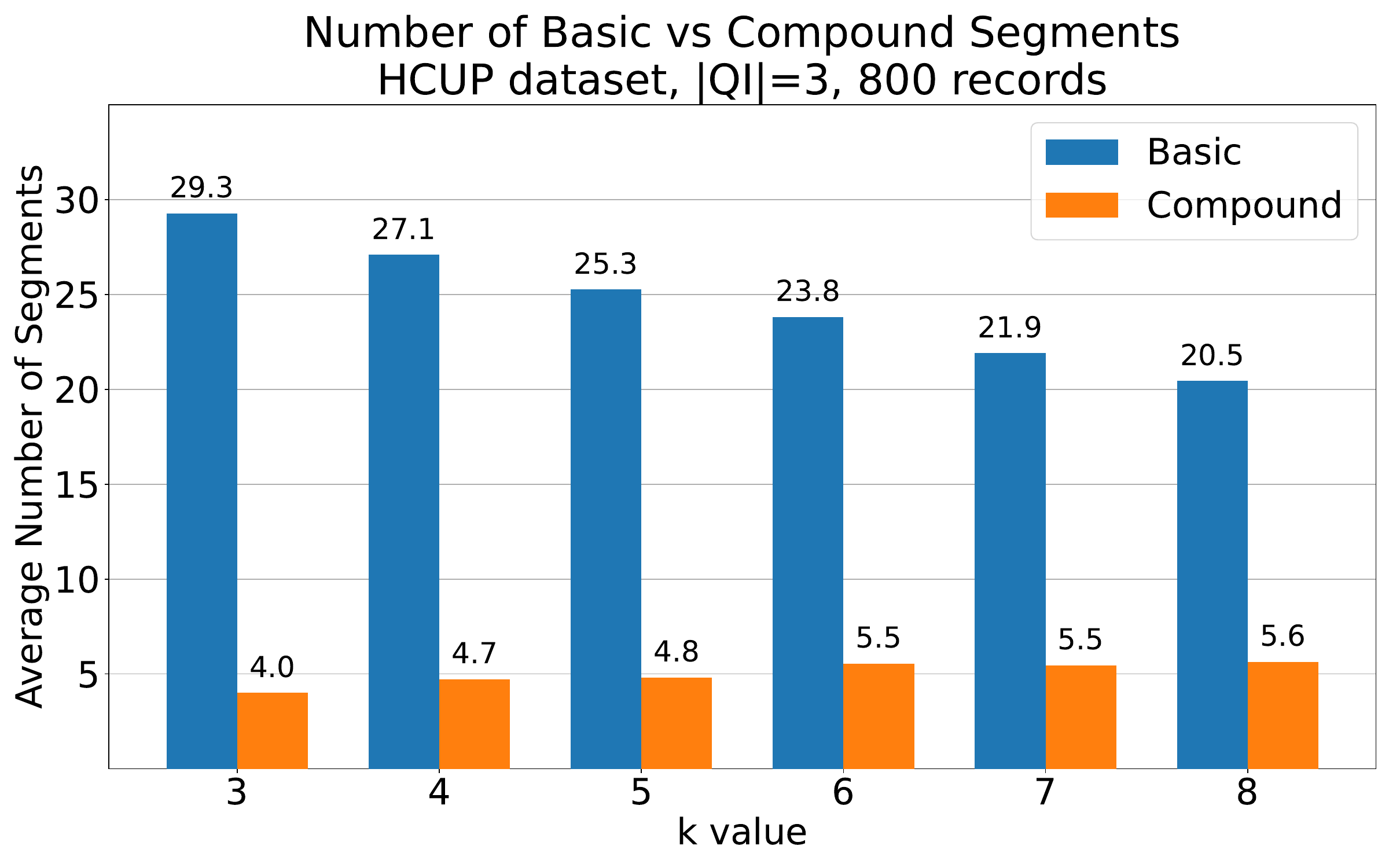}
        \label{fig:segments2_count}
    \end{subfigure}
    \vspace{-0.2cm}
    \caption{An analysis of the distribution of equivalence classes corresponding to basic segments versus those corresponding to compound segments, for varying values of $k$, on the \clinic and \texttt{HCUP} datasets. }
    \label{fig:segments_count}
\end{figure}

\subsection{CRA Runtime for Outliers}
\label{section:outlier_runtime}

\begin{table}[t]
\centering
\scriptsize
\resizebox{0.48\textwidth}{!}{%
\begin{tabular}{|c|c||c|c|c|c|c|}
\hline
\multirow{2}{*}{\textbf{Dataset}} & \multirow{2}{*}{\textbf{\# QIs}} & \multicolumn{5}{c|}{\textbf{Average runtime (seconds)}} \\
\cline{3-7}
& & $k=3$ & $k=4$ & $k=5$ & $k=6$ & $k=7$ \\
\hline
\hline
\multirow{3}{*}{\texttt{HCUP}} 
& $|QI|=2$ & $9.9\times{10^{-3}}$ & $3.5\times{10^{-3}}$ & $4.0\times{10^{-3}}$ & $4.4\times{10^{-3}}$ & $4.7\times{10^{-3}}$ \\
\cline{2-7}
& $|QI|=3$ & $1.8\times{10^{-1}}$ & $7.8\times{10^{-1}}$ & $4.0\times{10^{-1}}$ & $2.3\times{10^{-1}}$ & $6.6\times{10^{2}}$ \\
\cline{2-7}
& $|QI|=4$ & $6.6$ & $5.0\times{10^{3}}$ & $8.8\times{10^{3}}$ & $4.3\times{10^{4}}$ & $3.8\times{10^{4}}$ \\
\hline
\hline
\multirow{2}{*}{\texttt{\clinic}} 
& $|QI|=2$ & $2.6\times{10^{-2}}$ & $2.4\times{10^{-2}}$ & $5.5\times{10^{-2}}$ & $2.7\times{10^{-1}}$ & $1.7$ \\
\cline{2-7}
& $|QI|=3$ & $1.3$ & $2.3\times{10}$ & $3.3\times{10^3}$ & $6.2\times{10^3}$ & $2.7\times{10^4}$ \\
\hline
\end{tabular}%
}
\caption{Average CRA Runtime per outlier}
\label{table_runtime_outlier}
\end{table}
In Table~\ref{table_runtime_outlier}, we report the average runtime for CRA for outliers across different values of $k$ and number of quasi-identifiers. 
Unlike the CRA runtime for equivalence classes, the runtime for outliers does not increase with increasing $k$ values. 
In the case of equivalence classes, the number of variables associated with the linear programming formulation grows with $k$, since larger equivalence classes span more basic segments. 
However, this dependency does not apply to outliers. 
Outliers can only occupy basic segments not already assigned to equivalence classes.
Therefore, the number of basic segments associated with outliers is equal to the difference:
$(\text{total number of basic segments in the segment grid}) - (\text{number of basic segments 
assigned to equivalence classes})$.
As a result, the number of basic segments, and consequently the number of variables associated with Linear Programming, depends on the number of equivalence classes.
Since number of equivalence classes varies according to the dataset instead of the value of $k$, we don't see a monotonic increase in runtime with the values of $k$. 
On the other hand, a clear trend is visible when increasing the number of quasi-identifiers. 
This is expected since the number of basic segments in the segment grid grows with increase in the number of dimensions (or quasi-identifiers), leading to an increase in the number of variables associated with the linear programming formulation. 
This explains the more predictable runtime increase with an increase in number of quasi-identifiers.

\end{document}